\documentclass[11pt, a4paper]{article}
\usepackage[utf8x]{inputenc}
\usepackage{graphicx}
\usepackage{jcappub_ver}
\usepackage{wrapfig}

\hypersetup{pdftitle=CMB seen through random Swiss Cheese}

\newcommand{\be}{\begin{equation}} 
\newcommand{\ee}{\end{equation}}
\newcommand{\ba}{\begin{align}}
\newcommand{\ea}{\end{align}}
\newcommand{\bdm}{\begin{displaymath}} 
\newcommand{\edm}{\end{displaymath}}
\newcommand{\tthe}{\tilde{\theta}}
\newcommand{\tsig}{\tilde{\sigma}}
\newcommand{\thh}{\tilde{h}}
\newcommand{\bea}{\begin{eqnarray}} 
\newcommand{\eea}{\end{eqnarray}}
\newcommand{\mpc}{\ h^{-1}\mathrm{Mpc}}
\newcommand{\D}{\mathrm{d}}
\newcommand{\re}[1]{(\ref{#1})}
\renewcommand{\l}{\lambda}

\newcommand{\ie}{i.e.\ }

\begin{document}

\title{CMB seen through random Swiss Cheese}

\author{Mikko Lavinto and}
\author{Syksy R\"as\"anen}

\affiliation{Physics Department, University of Helsinki and Helsinki Institute of Physics \\
P.O. Box 64, FIN-00014, University of Helsinki, Finland} 

\emailAdd{mikko.lavinto@helsinki.fi}
\emailAdd{syksy.rasanen@iki.fi}

\abstract{
We consider a Swiss Cheese model with a random arrangement of 
Lema\^itre--Tolman--Bondi holes in $\Lambda$CDM cheese.
We study two kinds of holes with radius $r_b=50 \mpc$,
with either an underdense or an overdense centre, called the
open and closed case, respectively.
We calculate the effect of the holes on the temperature,
angular diameter distance and, for the first time in Swiss Cheese
models, shear of the CMB.
We quantify the systematic shift of the mean and the statistical
scatter, and calculate the power spectra.

In the open case, the temperature power spectrum is three orders
of magnitude below the linear ISW spectrum. It is sensitive to
the details of the hole, in the closed case the amplitude 
is two orders of magnitude smaller.
In contrast, the power spectra of the distance and shear are more
robust, and agree with perturbation theory and previous Swiss Cheese
results.
We do not find a statistically significant mean shift in the
sky average of the angular diameter distance, and obtain the
95\% limit $|\Delta D_A/\bar{D}_A|\lesssim10^{-4}$.

We consider the argument that areas of spherical surfaces are
nearly unaffected by perturbations, which is often invoked in light
propagation calculations. The closed case is consistent with this
at 1$\sigma$, whereas in the open case the probability is only 1.4\%.
}

\maketitle

\section{Introduction}

\paragraph{Inhomogeneities, mean quantities and sampling in light propagation.}

Studies of the non-linear effect of inhomogeneities
on light propagation go back to at least the proposal by
Zel'dovich in 1964 that, due to clumping of matter,
lines of sight could pass through regions of the universe
that are more underdense than the mean,
changing the observed distance-redshift relation \cite{Zeldovich:1964}.
The work that followed considered a modification to the
optical equations such that light rays see only
a constant fraction of the density of a Friedmann--Robertson--Walker
(FRW) universe
\cite{Dashevskii:1965, Dashevskii:1966, Bertotti:1966, Gunn:1967},
now referred to as the Dyer--Roeder formula
\cite{Dyer:1972, Dyer:1973zz, 1974ApJ...189..167D, 1975ApJ...196..671R}.
The applicability  of the formalism has since been much studied
\cite{Ehlers:1986, Futamase:1989, Kasai:1990, Vanderveld:2008, Rasanen:2008be,  Okamura:2009, Rasanen:2009, Szybka:2010, Clarkson:2011br, Bolejko:2012ue, Fleury:2014},
and it has been generalised by making the density sampling factor
redshift-dependent and by changing the expansion rate along the light ray
\cite{Linder:1988a, Linder:1988b, Linder:1988c, Linder:1997, Linder:1998, Kantowski:2003, Mattsson:2007tj, Bolejko:2010nh, Clarkson:2011br, Busti:2013}.
The effect of inhomogeneities has been studied also by a variety of other
methods; for discussion and references, see e.g. \cite{Rasanen:2008it}.

It is sometimes claimed that even though inhomogeneities can change
the distance-redshift relation for some lines of sight, on average
it will not deviate from the FRW result.
This was argued by Weinberg in 1976 \cite{Weinberg:1976},
on the basis that because inhomogeneities do not change
the number of photons, the total flux through a spherical surface
would be conserved.
It was already pointed out by Bertotti in 1966 \cite{Bertotti:1966}
that this argument is wrong, because it assumes that the area of
the sphere is unchanged by inhomogeneities,
which, however, is precisely the question under study.
In general, the area of a sphere can be strongly affected by
inhomogeneities, and explicit examples have been presented
\cite{Mustapha:1997, Ellis:1998ha, Ellis:1998qga},
including a case where the metric perturbations around the
FRW universe are small \cite{Enqvist:2009} (see also
\cite{Mustapha:1998, Enqvist:2007, GarciaBellido:2008, Bolejko:2011b, Sundell:2015}).
The redshift can also change significantly, modifying the
distance-redshift relation \cite{Lavinto:2013}.

It seems that if structures are small compared to the distance
travelled by the light and their distribution is statistically
homogeneous and isotropic and evolves slowly
compared to the time it takes for light to cross them
--in short, if light rays sample the structures fairly--
then the change in the redshift and distance can be expressed
in terms of the average expansion rate
\cite{Rasanen:2008it, Rasanen:2008be, Rasanen:2009, Bull:2012, Lavinto:2013}.
Thus, if the average expansion rate is close to the FRW case,
this is expected to be true also for the redshift and the distance.
(It is known that if the metric is close to FRW,
so are the redshift and the average expansion rate \cite{Rasanen:2011b}.)

Attention has thus focused either on studying the possibility of a large
effect of structures on the average expansion rate, \ie backreaction
\cite{Buchert:2011sx, Rasanen:2011a}, or analysing the small
deviations in the case when the average expansion rate is close to FRW.
A useful treatment of the latter situation is given by Swiss Cheese
models, where inhomogeneities are modelled by cutting
holes into a spacetime (called the background, or cheese)
and replacing them with patches of another spacetime. This was
originally done for an Einstein--de Sitter background and Schwarzschild
holes \cite{Einstein:1945id}. However, the setup can be easily generalised
to any FRW background and any Szekeres solution \cite{Szekeres:1974},
provided the Darmois junction conditions are satisfied
\cite{Darmois1927, Israel:1966, Bonnor:1981, Carrera:2008}.
If the Szekeres holes are small, long-lived and non-singular
(including the absence of surface layers), then the average
expansion rate is close to (but not exactly the same as)
that of the background \cite{Lavinto:2013}.
A Swiss Cheese model can be statistically homogeneous and isotropic,
if the holes are small and their distribution is chosen appropriately.
As Swiss Cheese is an exact solution of the Einstein equation,
no perturbative or Newtonian approximations are needed.
The model is rather tractable in
the sense that the solution to the Einstein equation can be expressed
in terms of elliptic integrals, and in the case when cosmological
constant is zero, it can be written parametrically in terms of
elementary functions \cite{Plebanski:2006sd}.
It is also possible to treat more complicated structures than in those exact
solutions in which there is no FRW background and the sources are discrete
\cite{Clifton:2009b, Clifton:2009c, Clifton:2011, Liu:2015, Sanghai:2015}
(see also \cite{Clifton:2012, Yoo:2012, Bruneton:2012a, Bentivegna:2012, Bruneton:2012b, Larena:2012, Bruneton:2013, Bentivegna:2013, Korzynski:2013, Korzynski:2014, Korzynski:2015}).

Light propagation in Swiss Cheese models has been extensively studied
\cite{Kantowski:1969, Kantowski:1995, Kantowski:1998, Sugiura:1999b, Claudel:2000, Kantowski:2000, Brouzakis:2006, Biswas:2007gi, Brouzakis:2007, Marra:2007pm, Marra:2007gc, Bolejko:2008wx, Bolejko:2008xh, Vanderveld:2008, Marra:2008, Ghassemi:2009, Clifton:2009a, Kostov:2009, Kostov:2010, Bolejko:2010eb, Bolejko:2010c, Szybka:2010, Marra:2011, Clarkson:2011br, deLavallaz:2011, Flanagan:2011, Flanagan:2012, Bolejko:2012, Fleury:2013sna, Fleury:2013uqa, Fleury:2014, Peel:2014, Koksbang:2015jba}.
In particular, Swiss Cheese models have been used to study fluctuations
in redshift and distance of the cosmic microwave background (CMB)
\cite{Bolejko:2008xh, Valkenburg:2009, Bolejko:2011a}.
Some papers, including recent ones, based either on Swiss Cheese and
perturbative calculations, have claimed surprisingly large
effects on the distance, given that the average expansion rate
has been close to FRW
\cite{Marra:2007pm, Marra:2008, Bolejko:2011a, Fleury:2013sna, Fleury:2013uqa, Fleury:2014, Clarkson:2014pda}.
Such deviations have been due to selection effects, some of
which are quite clear and others rather subtle.

We investigate the systematic shift and statistical fluctuations in
the redshift, angular diameter distance and shear
of the CMB in a Swiss Cheese model.
We consider a fully randomized hole distribution in a single
fixed spacetime, to avoid biased sampling.
This is the first Swiss Cheese CMB calculation of the distance
fluctuations with randomised holes in a $\Lambda$CDM background.
It is also the first study of the CMB shear in a Swiss Cheese model.
We use the Lema\^itre--Tolman--Bondi (LTB)
\cite{Lemaitre, Tolman, Bondi, Plebanski:2006sd}
solution for the holes, and do the calculation
for two different density profiles, one with a
central overdensity and another with a central underdensity, for comparison.
We calculate the sky maps and power spectra for large scales
($l\lesssim100$) for the perturbations in redshift, angular
diameter distance and integrated null shear generated
by the holes between the CMB last scattering surface and the observer.
We also study sky averages, including the average of the flux and its inverse.
In section 2 we define our model. In section 3 we go through the
light propagation calculation. In section 4 we present our results,
compare to previous work, with a focus on selection effects
and flux conservation arguments, and in section 5 we summarise.

\section{Model construction}

\paragraph{The Lema\^itre-Tolman-Bondi solution.}

The LTB solution \cite{Lemaitre, Tolman, Bondi, Plebanski:2006sd} is the spherically symmetric special case of the Szekeres dust solution. The line element is
\be
\D s^2=-\D t^2+\frac{{R'}^2(t,r)}{1+2E(r)} \D r^2 + R^2(t,r) \D \Omega^2 \ ,
\ee
where $E(r)$ quantifies the spatial curvature of the shell at coordinate radius $r$; we denote derivative with respect to $r$ by prime and derivative with respect to $t$ by dot. In units where $c = G = 1$, the Einstein equation reduces to the equation of motion for $R$,
\be
\dot{R}^2(t,r) = 2E(r) + \frac{2M(r)}{R(t,r)} + \frac{1}{3} \Lambda R^2(t,r) \ , \label{eq:r-eom}
\ee
where $\Lambda$ is the cosmological constant and $M(r)$ is a free function related to the dust density as
\be
\rho(t,r) = \frac{M'(r)}{4 \pi R^2(t,r) R'(t,r)} \ .
\ee

\noindent The solution to (\ref{eq:r-eom}) can be found by integrating
\be
t-t_B(r) = \int_0^{R(t,r)} \frac{\D R}{\sqrt{\frac{2M}{R}+2E+\frac{1}{3}\Lambda R^2}} \ ,
\ee
and numerically finding the root(s). This integral cannot
generally be expressed in terms of elementary functions. We solve it
by numerical integration and the algorithm presented in
\cite{Carlson:1995, Valkenburg:2012}. The bang time function $t_B(r)$
indicates the time of the initial singularity, $R(t_B(r),r) = 0$. 

The expansion rate is
\ba
H(t,r) &= \frac{1}{3}\left(\frac{\dot{R}'}{R'} + \frac{2\dot{R}}{R}\right) \ .
\end{align}

The local homogeneous FRW dust solution is a special case with
\be E = -\frac{1}{2} K r^2, \hspace{1cm} M = M_0 r^3, \hspace{1cm} t_B = 0 \ , \ee

\noindent where $K$ and $M_0$ are constants, leading to $R(t,r) = a(t)r$.

\paragraph{Junction conditions and shell crossings.}

Two LTB spacetimes can be joined together into a solution of the Einstein equation without surface layers if and only if the metric and the extrinsic curvature are continuous across the boundary. These are known as the Darmois junction conditions \cite{Darmois1927, Israel:1966}. In order to have a physically viable model, shell crossings also have to be avoided \cite{Hellaby:1985zz}. Since the LTB solution has only dust matter, there are no physical processes that can stop two nearby radial shells from overlapping if they expand at different rates. This leads to a divergence in density and a breakdown of the model. We construct the models so that there are no shell crossings between the time of the big bang and today. Requiring that the LTB holes are free of shell crossings, collapse singularities and surface layers, and that their radius today is much smaller than the Hubble scale, constrains them to have nearly the same volume, average density and total mass as the piece of cheese they displace \cite{Lavinto:2013}.

\paragraph{Choice of functions.}

\begin{figure}
\begin{center}
\includegraphics[width=0.5\linewidth]{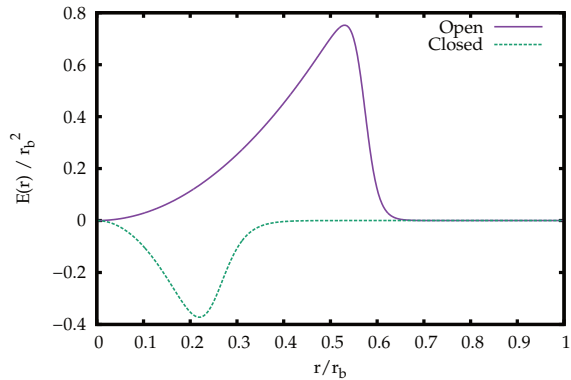} 
\caption{The function $E(r)$ in units of $r_b^2$.}
\label{fig:curv}
\end{center}
\end{figure}

The three free functions $E(r)$, $M(r)$ and $t_B(r)$ specify the model completely. Models with $t_B'\neq0$ contain decaying modes \cite{Silk:1977}, which would correspond to the universe being far from homogeneous and isotropic at early times. We consider only models where $t_B$ is constant; without loss of generality we then take $t_B=0$. Given the junction conditions, which fix the value of $M(r)$ on the embedding surface and the assumption that $M(r)$ is monotonic, its functional form is completely degenerate with the definition of the radial coordinate $r$. Our choice of  $M(r)$ and $E(r)$ is given below, along with $t_B(r)$:
\ba
M(r) &= M_0 r^3 \\
E(r) &= E_0 r^2 \left[1-\tanh\left(\frac{r-r_t}{r_\sigma}\right)\right] \\
t_B(r) &= 0 \ ,
\end{align}
where $M_0$ and $E_0$ are constants. The junction conditions fix $M_0 = \frac{4 \pi r_b^3}{3} \bar{\rho}(t_0)$, where bar denotes background quantity and $t_0$ is the present time. The form of $E(r)$, shown in figure \ref{fig:curv}, has been chosen because of its smoothness, which makes it possible to speed up the calculation by evaluating $R$ using cubic splines. The constant $E_0$ determines the sign and magnitude of the spatial curvature in the hole. Parameters $r_t$ and $r_\sigma$ are the transition radius and width, respectively, where the profile changes smoothly from spatially curved to flat. We choose these constants so that at the embedding boundary $E(r)$ is zero within numerical precision and therefore the Darmois junction conditions are satisfied to good precision, while keeping the model free of shell crossings and also making sure it is very non-linear today; the values are given in table \ref{table:params}. The fact that the
Darmois conditions are not exactly satisfied is not a problem. We could make the matching exact by multiplying $E(r)$ with a function that is zero at the boundary and switches rapidly to unity away from it, without affecting the results.

\paragraph{Swiss Cheese.}

\begin{table}\begin{center}
\begin{tabular}{c|c c}
 & Open & Closed \\ 
\hline 
$r_b$ & $50\mpc $  & $50\mpc$\\
$r_t/r_b$ & $20/35$ & $5/20$ \\
$r_\sigma/r_b$ & $1/{35}$ & $1/{20}$ \\ 
$E_0$ & $1.41$ & $-5$ \\
$\langle \dot{\delta} \rangle_V(t_0)$ & $0.156 H_0$ & $0.010 H_0$ \\
$\langle \Delta H / \bar{H} \rangle_V(t_0)$ & $-2.0 \cdot 10^{-7}$ & $-1.8 \cdot 10^{-9}$ \\
\end{tabular} 
\caption{Parameters of the two models: the embedding radius $r_b$, transition radius $r_t$ and transition width $r_\sigma$ and the magnitude of spatial curvature $E_0$. Also listed are the proper volume averages, denoted by $\langle\rangle_V$, of $\dot{\delta}$ and $\Delta H / \bar{H}$ today.}
\label{table:params}
\end{center}
\end{table}

\begin{figure}
\includegraphics[width=0.49\linewidth]{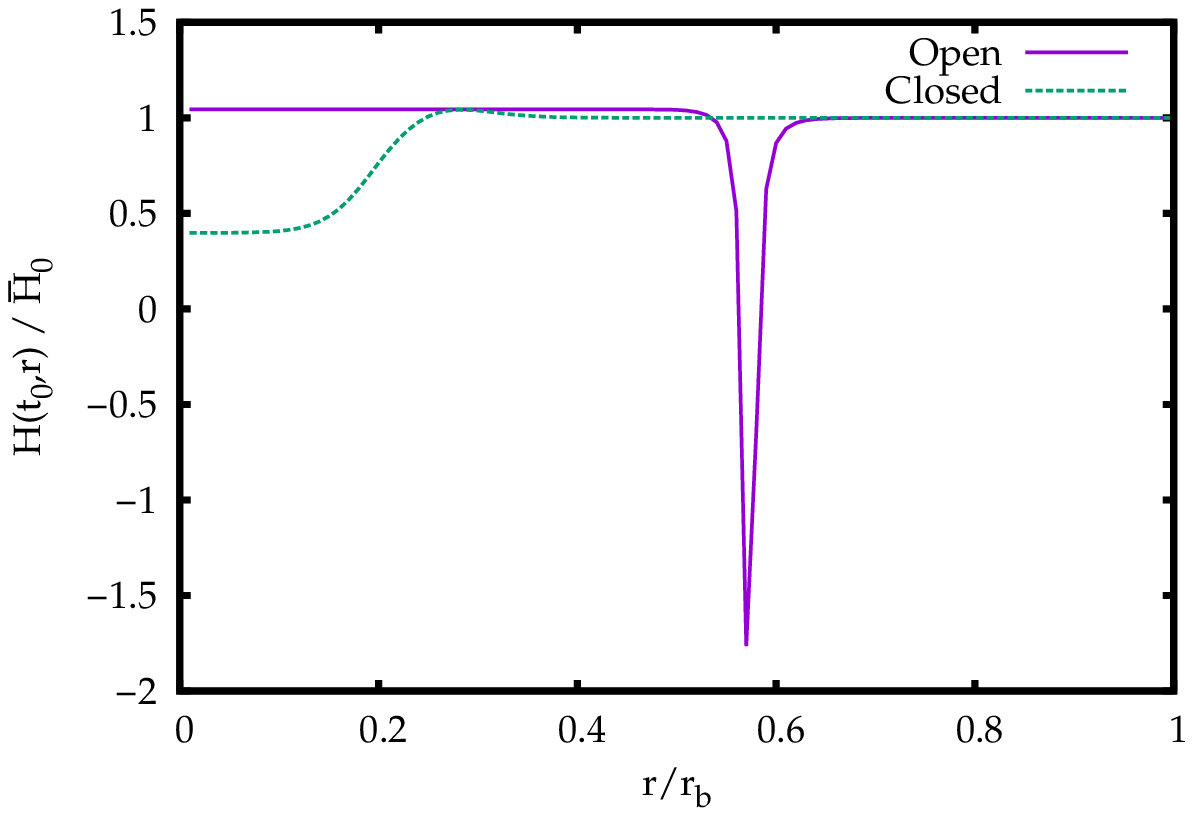} 
\includegraphics[width=0.49\linewidth]{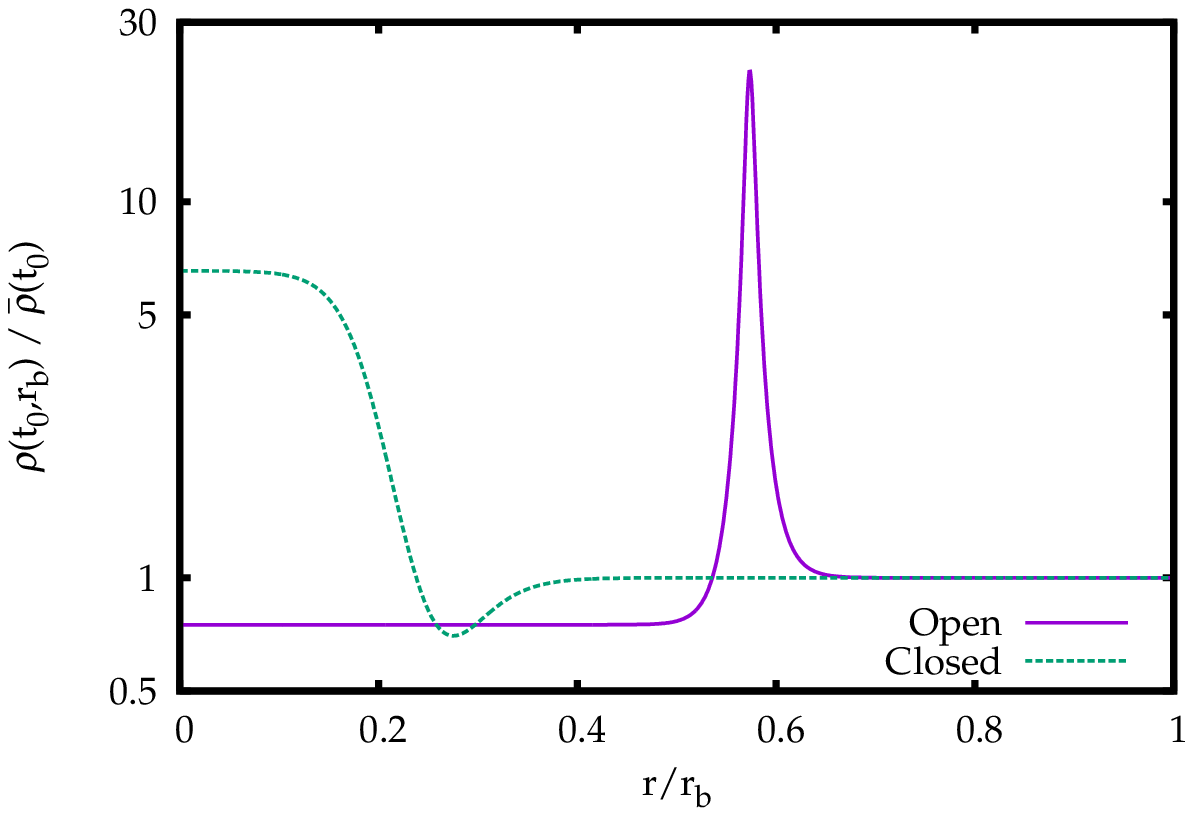} 
\caption{The expansion rate (left) and density (right) of the holes normalised to the background today, as a function of $r/r_b$. Note that the density plot is logarithmic.}
\label{fig:expansion}
\end{figure}

We take the background to be a spatially flat FRW spacetime with dust
and a cosmological constant, with present day density parameters
$\Omega_\Lambda=0.7$, $\Omega_m=0.3$. We consider two kinds of
LTB holes, called open and closed, to see how the
results depend on the details of the hole.
In the open case, there is an underdense void in the centre,
surrounded by a thin overdense shell.
The closed case has an overdense centre surrounded by an underdensity.
The spatial curvature closely follows the density profile, so it 
is negative at the location of the underdensity and positive at the
overdensity. 
Despite the names, the average spatial curvature over one hole
is positive in both cases, although very small,
$\langle^{(3)} R\rangle_V/(6 \bar{H}_0^2)=5.7\cdot10^{-8}$
in the open model and $1.3\cdot10^{-8}$ in the closed model,
where $\langle\rangle_V$ denotes proper volume average and
$\bar{H_0}$ is the background Hubble rate today.

In both cases, the holes are embedded in the background on a
comoving boundary with coordinate radius $r_b = 50 \mpc$.
However, the profiles are close to the background already far from
the boundary, so the size of the inhomogeneities is smaller than the
embedding radius, approximately $30 \mpc$ for the open model and $20 \mpc$
for the closed model, as shown in figure \ref{fig:expansion}.
In order to avoid shell crossings, there is a trade-off between the
non-linearity of the profile and distance of the inhomogeneous
features from the boundary, and we have constructed the holes to be clearly
in the non-linear regime today.

The energy density in our model does not have a radiation component,
so we are missing the early integrated Sachs--Wolfe (ISW)
effect. In the real universe,
the contribution of radiation to the energy density
at the last scattering surface is about 25\%. However, we are in any case not
interested in modelling the effects at early times, when the holes
are in the linear regime. Therefore, we position the sources
(as well as the observer) in the cheese. We thus also do not have
an ordinary Sachs-Wolfe effect, nor monopole or dipole terms
arising from the observer's location (the dipole is generally
the dominant contribution; see e.g. \cite{Nadathur:2014}).
The signal is generated only by secondary effects as the photons
propagate through the universe.

We arrange the holes randomly around the observer, with the locations
drawn from a uniform distribution and overlapping holes removed.
Although this procedure generates some correlation between voids,
their distribution is close to uniformly random. The fraction of
the total volume in holes, the packing fraction, is $0.34$,
close to the packing efficiency of 0.35 in \cite{Kostov:2009, Kostov:2010}.
The effective packing fraction, defined as the fraction of volume
in regions that differ significantly from FRW, is smaller, as the profiles
tend to the FRW case far from the boundaries, as noted above.
We use the same single arrangement of holes for both models and for all
geodesics, so correlations due to different rays having passed through
the same holes are included.
In contrast, most earlier works (\cite{Kostov:2009, Kostov:2010}
being a notable exception) have considered a hole distribution
that is dynamically generated along each beam separately
or treated statistically
\cite{Kantowski:1969, Kantowski:1995, Kantowski:1998, Sugiura:1999b, Kantowski:2000, Brouzakis:2007, Bolejko:2008xh, Vanderveld:2008, Szybka:2010, Bolejko:2011a, Flanagan:2011, Flanagan:2012, Fleury:2014, Koksbang:2015jba}
(see also
\cite{Holz:1997, Holz:1998, Kainulainen:2009a, Kainulainen:2009b, Kainulainen:2009c, Kainulainen:2010a, Kainulainen:2010b, Kainulainen:2011}),
or have arranged the holes straight on the line of sight
\cite{Brouzakis:2006, Marra:2007pm, Marra:2007gc, Clifton:2009a, Bolejko:2010eb, Bolejko:2010c, Marra:2011, Bolejko:2011a}
or in a regular lattice
\cite{Valkenburg:2009, Kostov:2009, Kostov:2010, Fleury:2013sna}.

In the real universe, there are correlated structures of various sizes.
It would be possible to construct a model with holes of different
radii, profiles and correlated locations to try to match the
matter power spectrum. The packing fraction could be increased
by using holes of different sizes.
By using holes that have a homogeneous region, it is also possible
to pack holes inside holes \cite{Korzynski:2014}.
However, as the holes in Swiss Cheese cannot overlap, merge
nor otherwise affect each other gravitationally, its ability to
model realistic large scale structure is limited.
We restrict ourselves to single size holes with an (almost) uniform
distribution.

\section{Light propagation}

\paragraph{The setup.} 

As long as the wavelength and the wavefront curvature radius are much
smaller than spacetime curvature radius, light propagation can be described
in the geometrical optics approximation, where light travels as plane
waves on null geodesics \cite{P.Schneider1992} (p. 93). Let a null
geodesic $x^\mu(\lambda)$ be parametrized by an affine parameter
$\lambda$. Its tangent vector $k^\mu = \frac{\D x^\mu}{\D \lambda}$
satisfies the null condition $k_\mu k^\mu = 0$ and the geodesic
equation $k^\mu \nabla_{\mu} k^\nu = 0$. If the wave vector is given
in units where the photon energy at observation is equal to $1$, then
the redshift measured by an observer with four-velocity $u^\mu$ is
given by $1+z = -u_\mu k^\mu$. We consider observers that comove with
the dust, so $z = k^t-1$. The temperature is related to the redshift as
$T\propto 1+z$.

We start the light rays from the observer and follow the evolution
of a bundle of null geodesics back in time 
to the source, using the Sachs formalism to
calculate the angular diameter distance and the null shear.
We continue the integration until the background redshift reaches 1090,
corresponding to the last scattering surface.
This choice of direction makes the calculation a lot simpler,
and is natural in the sense that all observed geodesics
converge at the observer today.
However, in reality photons travel from the source towards the
observer, so it might seem that this standard procedure solves
the wrong physical problem.
For the distance, the resolution is well known:
backwards integration gives the angular diameter distance $D_A$,
whereas forwards integration gives the luminosity distance $D_L$
divided by $1+z$ \cite{Ellis:1971pg}. As the distances are related
due to the Etherington reciprocity theorem \cite{Eth33,Ellis:1971pg}
as $D_L=(1+z)^2 D_A$, it is simple to correct for this.
In appendix \ref{sec:proof}, we prove a similar result for the shear:
as long as the integrated null shear is small, it does not depend on
whether the initial conditions are set at the observer or the source.

\paragraph{Geodesic equations and redshift.}

Without loss of generality, we choose the LTB angular coordinates in each hole separately such that the light ray travels on the equator, $\theta = \pi/2$. Then the null geodesic equations reduce to
\ba
\frac{\D \theta}{\D\lambda} &= 0 \\
\frac{\D \phi}{\D\lambda} &= \frac{c_\phi}{R^2} \\
\frac{\D z}{\D\lambda} &= -\frac{\dot{R}'}{R'} (1+z)^2+\frac{c_\phi^2}{R^2}\left(\frac{\dot{R}'}{R'} - \frac{\dot{R}}{R}\right)  \label{z-geod}\\
\frac{\D ^2 r}{\D \lambda} &= -2\frac{\dot{R}'}{R'}(1+z)\frac{\D r}{\D \lambda} - \left[ (1+2E) \frac{R''}{R} - E' \right] \left(\frac{\D r}{\D \lambda}\right)^2 + (1+2E)\frac{c_\phi^2}{R^3 R'} \ ,
\end{align}
where $c_\phi$ is an integration constant related to the impact parameter. 
The initial condition for the radial component of the tangent vector can be found using the null condition $k_\mu k^\mu = 0$,
\be \frac{\D r}{\D \lambda} = \frac{1}{R'} \sqrt{(1+z)^2 - \frac{c_\phi^2}{R^2}} \ . \ee

\paragraph{Distance and null shear.}

The angular diameter distance from observer $O$ to source $S$ is defined as 
\be D_A \equiv \sqrt{\frac{\delta A_S}{\delta \Omega_O}} \ , \ee
where $\delta A_S$ is the surface area of the source and $\delta \Omega_O$ is the solid angle in which the source is seen by the observer. We state our results in terms of $D_A$, which is related to the luminosity distance due to the Etherington reciprocity theorem as
\be D_L = (1+z)^2 D_A \ . \label{eq:reci} \ee

The evolution of the beam cross section $A$ is given by the Sachs optical equations. The area expansion rate of the beam is
\be \tthe \equiv \nabla_\mu k^\mu = \frac{1}{A} \frac{\D A}{\D \lambda} \ , \ee
and the null shear tensor projected onto a hypersurface orthogonal to the wave vector $k^\mu$ with projection tensor $\thh^{\alpha}_{\ \mu}$ is
\be
\tsig_{\mu \nu} \equiv \thh^{\alpha}_{\ \mu} \thh^{\beta}_{\ \nu} \nabla_{(\alpha} k_{\beta)} - \frac{1}{2} \theta \thh_{\mu \nu} 
\ee
\be \tsig^2 \equiv \frac{1}{2} \tsig_{\mu \nu} \tsig^{\mu \nu} = (\nabla_\mu k_\nu) (\nabla^\mu k^\nu) - \tthe^2 \ . \ee
The null shear tensor has two real degrees of freedom, and it can be written as
\be \tsig^A_{\ B} = \left( \begin{matrix} \tsig_1 & \tsig_2 \\ \tsig_2 & -\tsig_1 \end{matrix} \right) \ , \ee

\noindent where $A, B$ label directions along the lens plane
defined by $\thh_{\alpha\beta}$, and $\tsig = \sqrt{\tsig_1^2 + \tsig_2^2}$.

The Sachs equations are
\begin{align} 
\frac{\D \tthe}{\D \lambda} + \frac{1}{2} \tthe^2 + 2 \tsig^2 &= -R_{\mu \nu} k^\mu k^\nu \label{sachs-exp}  \\
\thh^{\alpha}_{\ \mu} \thh^{\beta}_{\ \nu} \frac{\D \tsig_{\alpha \beta}}{\D \lambda} + \tthe \tsig_{\mu \nu} &= - C_{\gamma \alpha \delta \beta} k^\gamma k^\delta \thh^{\alpha}_{\ \mu} \thh^{\beta}_{\ \nu} \ , \label{sachs-shear}
\end{align}
where $R_{\mu\nu}$ is the Ricci tensor and $C_{\gamma \alpha \delta \beta}$
is the Weyl tensor. In terms of the angular diameter distance,
(\ref{sachs-exp}) reads
\be 
\frac{\D^2 D_A}{\D \lambda^2} + \left(\tsig^2 + \frac{1}{2} R_{\mu \nu} k^\mu k^\nu\right)D_A = 0 \ . \label{sachs-d}
\ee
Due to the spherical symmetry of individual LTB solutions, (\ref{sachs-shear}) and (\ref{sachs-d}) can be simplified to
\ba
\frac{\D^2 D_A}{\D \lambda^2} + &\left[\tsig^2 + 4 \pi (1+z)^2 \rho \right]D_A = 0 \\ 
\frac{\D \tsig_1}{\D \lambda} + \tthe \tsig_1 &= \frac{c_\phi}{R^2} \left(4 \pi \rho - 3 \frac{M}{R^3}\right) \cos\psi \\
\frac{\D \tsig_2}{\D \lambda} + \tthe \tsig_2 &= \frac{c_\phi}{R^2} \left(4 \pi \rho - 3 \frac{M}{R^3}\right) \sin\psi \ ,
\end{align}
where $\psi$ is the angle between the line formed by intersection of the lens
plane defined by $\thh_{\alpha\beta}$ with the equator of the first LTB hole,
and the line formed by its intersection with the equator of the hole that
the beam is going through, so $\psi=0$ for the first hole.
The cosmological constant $\Lambda$ does not enter into the source terms, as it only contributes via terms proportional to $g_{\mu \nu}$, and $k^\mu$ is null. We solve these equations backwards in time starting from the observer with the initial conditions
\be \left. D_A \right|_{O} = 0, \hspace{0.5cm} \left. \frac{\D D_A}{\D \lambda}\right|_{O} = -H_0^{-1}, \hspace{0.5cm} \left.\tsig_n\right|_{O} = 0 \ \label{init} .\ee

Shear has been correctly treated in Swiss Cheese models only rarely
\cite{Szybka:2010, Lavinto:2013}. Usually that is not important,
because often only the distance is considered, and the effect of the
null shear on the distance is generally subdominant to the effect
of the density. The shear is also small in our case.

The covariant quantities $\tthe$ and $\tsig_n$ can be translated into the perturbative lensing formalism, where the relevant quantity is the amplification matrix
\cite{P.Schneider1992, Bartelmann:1999yn, Lewis:2006, Hanson:2009, Clarkson:2011br}
(see \cite{Clarkson:2015pia} for the second order corrections)
\be \label{am}
  \mathcal{A}^A_{\ B} = \left( \begin{matrix} 1-\kappa-\gamma_1 & \gamma_2 \\ \gamma_2 & 1-\kappa+\gamma_1 \end{matrix} \right)
\ee
\noindent that relates the null geodesic direction at the source to the
direction at the observer (see \cite{Fanizza:2014} for discussion of the LTB case).
The convergence $\kappa$ corresponds to the
change in the bundle area and the integrated null shear $\gamma$ gives the
deformation of the source image. The magnification
\bea \label{mu}
  \mu = \frac{\bar{D}_A^2}{D_A^2} = \det(\mathcal{A})^{-1} = \left[(1-\kappa)^2 - \gamma^2\right]^{-1}
\eea

\noindent gives the change in the source luminosity relative to the FRW
background, with $\gamma \equiv \sqrt{\gamma_1^2 + \gamma_2^2}$.

In the weak lensing limit, $|\kappa| \ll 1$ and $|\gamma_n| \ll 1$, we have
\be \label{mud}
  \mu = \frac{\bar{D}_A^2}{D_A^2} \simeq 1 + 2\kappa \ , \hspace{1cm}
\gamma_i \simeq \int \D \lambda \tsig_i \ ,
\ee
where $\bar{D}_A$ is the angular diameter distance in the background FRW universe, so
\be \frac{\Delta D_A}{\bar{D}_A} \equiv \frac{D_A - \bar{D}_A}{\bar{D}_A} \simeq - \kappa \ . \ee
Another common way to describe lensing is to use the lensing potential $\psi$, defined via
\be A_{A B} = \delta_{A B} - \partial_A \partial_B \psi \ , \ee 
for example the Planck lensing results are given in terms of $\psi$ \cite{Planck:Lens}. For more on the relation of the different measures of lensing, see \cite{Schneider:2010, Clarkson:2011br}.

It is convenient to use the integrated null shear $\gamma$ instead of $\tsig$, as $\gamma$ describes the cumulative shearing along the beam path and is independent of the propagation direction (at least for small $\gamma$, see appendix \ref{sec:proof}). We give our results in terms of $\Delta T/\bar{T} \equiv (T - \bar{T})/\bar{T}$, $\Delta D_A/\bar{D}_A$ and $\gamma$.

\section{Results and discussion}

\paragraph{Single hole.}

\begin{figure}
\includegraphics[width=0.49\linewidth]{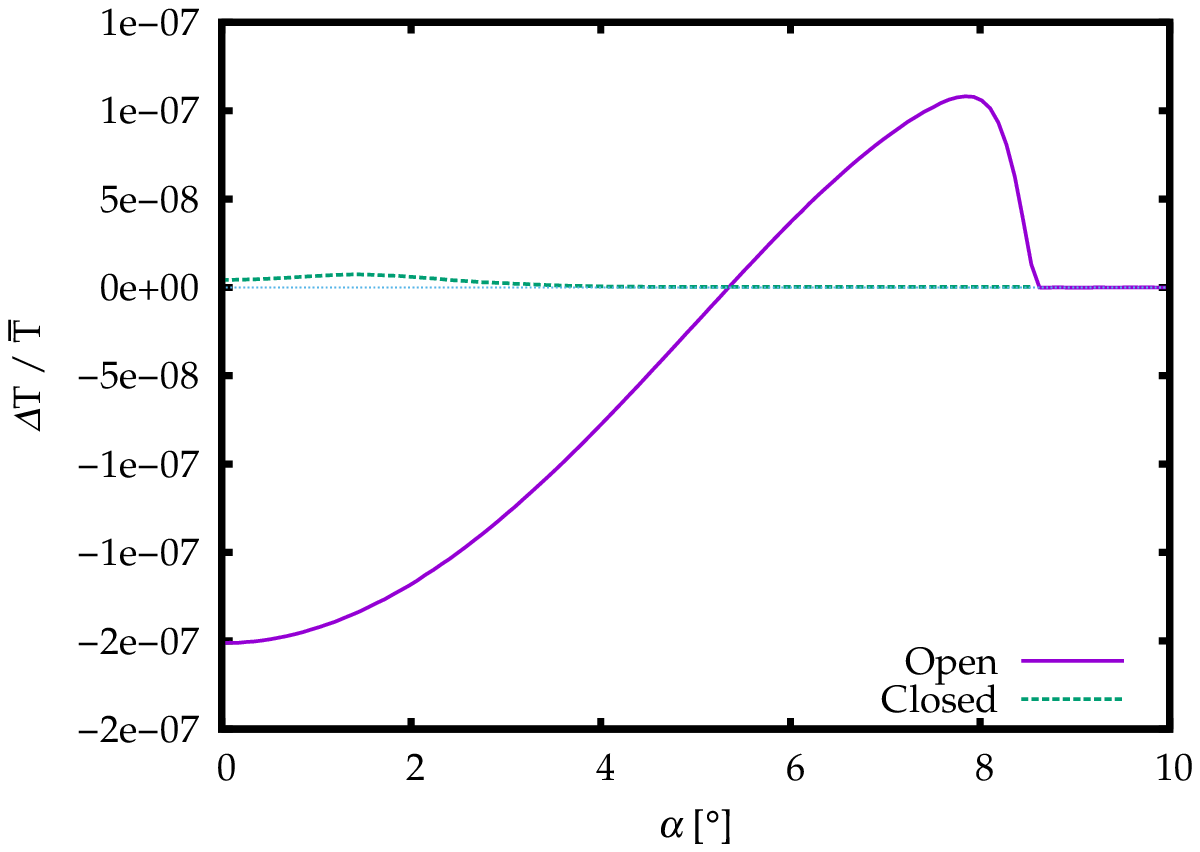} 
\includegraphics[width=0.49\linewidth]{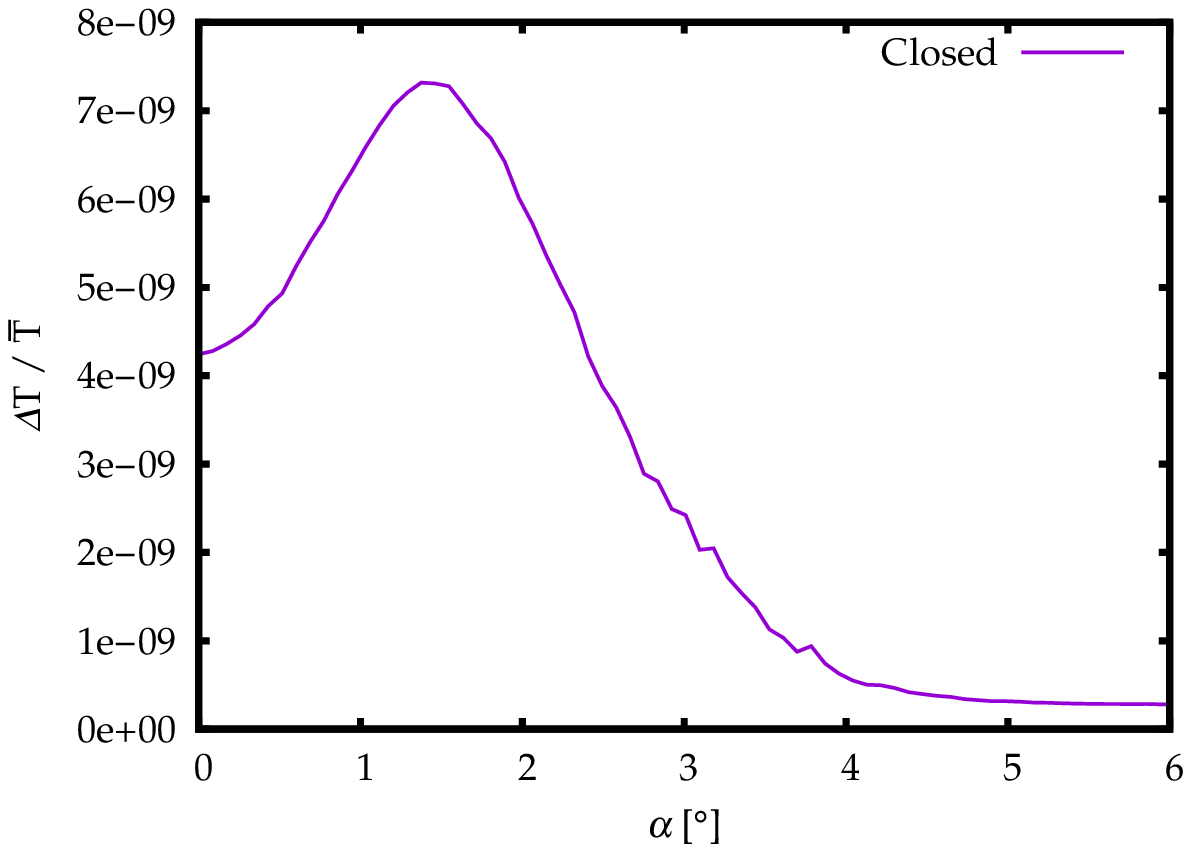} 
\caption{The angular profile of $\Delta T/\bar{T}$ for the open and closed
models (left) as a function of the viewing angle $\alpha$ for a single hole,
with its centre located at a coordinate distance of $200 \mpc$.
The result for the closed model is so much smaller
that it is also plotted separately (right).}
\label{fig:Tprofiles}
\end{figure}

\begin{figure}
\includegraphics[width=0.49\linewidth]{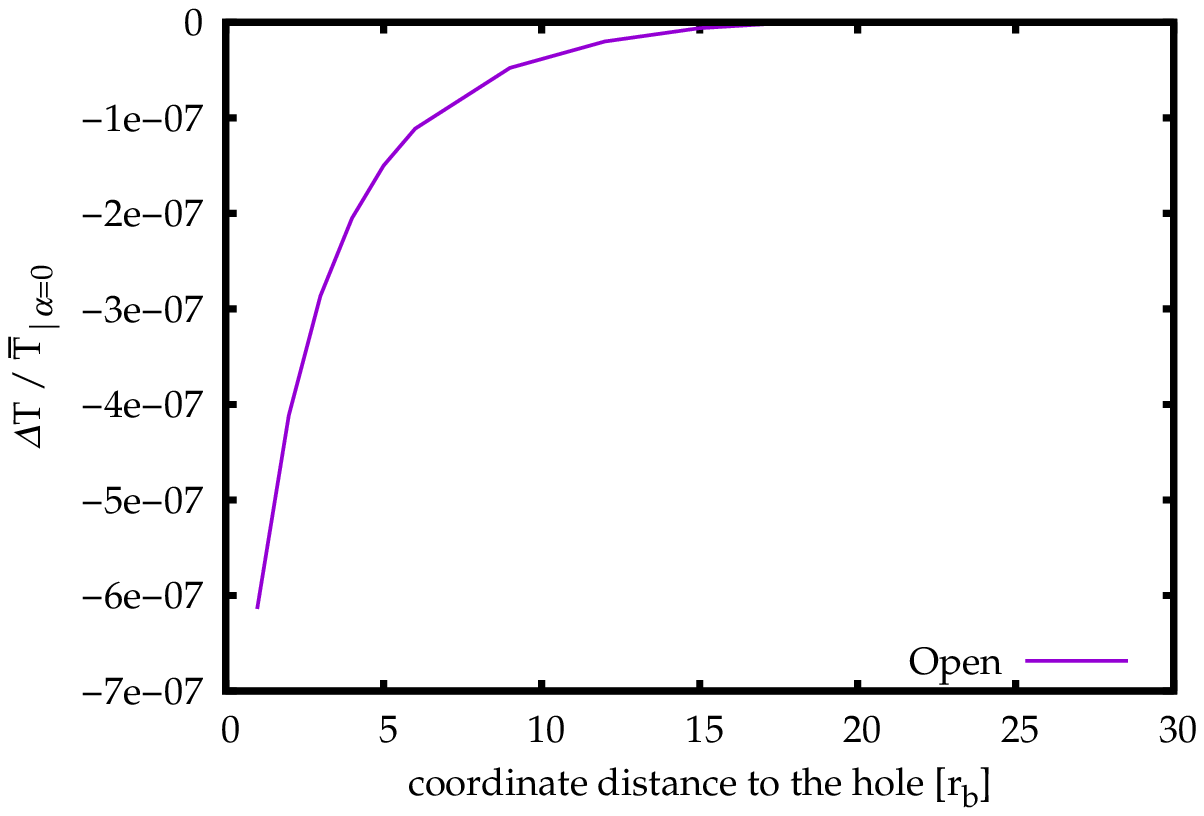} 
\includegraphics[width=0.49\linewidth]{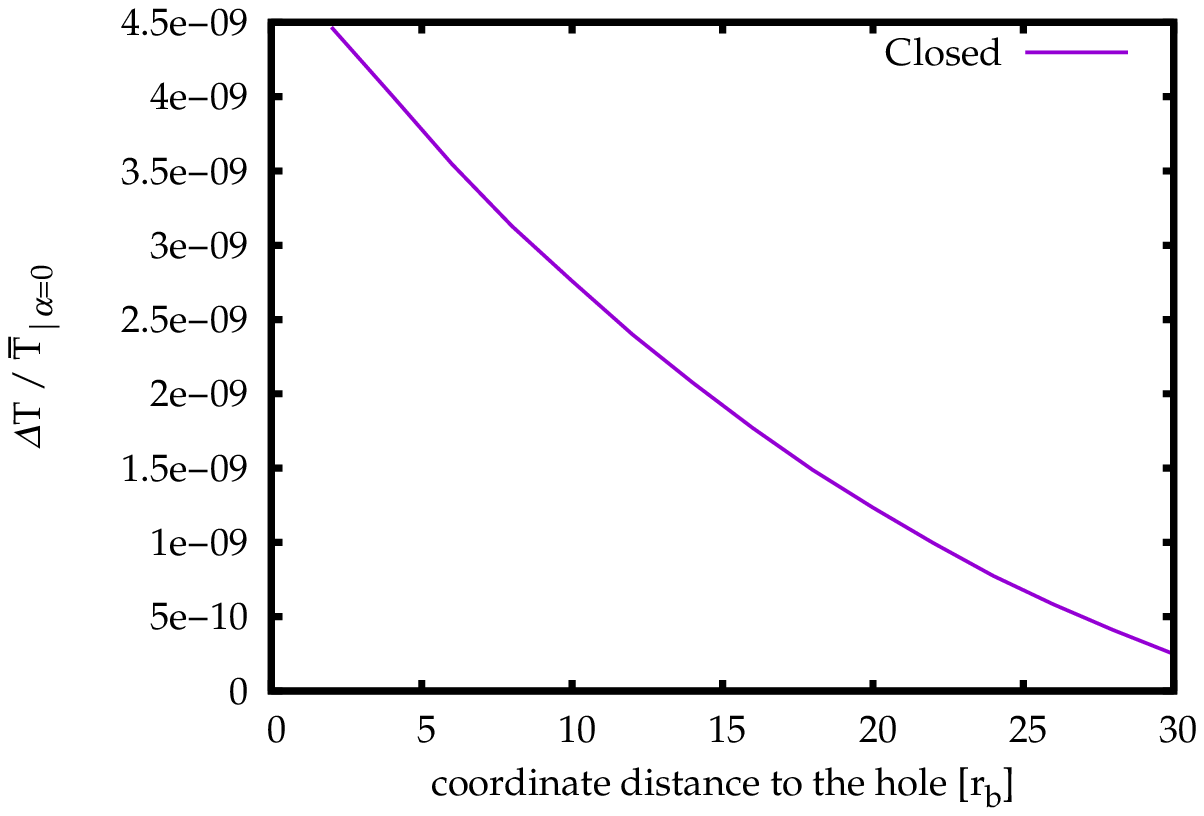}
\caption{Temperature shift for a radial light ray as a function of the distance to the centre in units of $r_b$, in the open (left) and closed (right) case.}
\label{fig:Tdistance}
\end{figure}

Let us first consider the effect of a single hole on the temperature,
distance and null shear.
In the open model, the temperature perturbation is negative in the
centre, surrounded by a positive ring, corresponding
to the underdense centre and the overdense shell, respectively.
For the closed model, the temperature perturbation is everywhere positive.
The temperature profiles are shown in figure \ref{fig:Tprofiles}.
The maximum amplitude of $\Delta T/\bar{T}$ for a hole at a distance
of 200$\mpc$ is $2\cdot10^{-7}$ in the open case and $7\cdot10^{-9}$
in the closed case, a ratio of 30.
In the open case, the amplitude decreases rapidly
with the distance to the hole, as shown in figure \ref{fig:Tdistance}.
For holes close to the observer, the maximum amplotude rises to $10^{-6}$.
In the closed case, the fall-off is less steep, because the proper
volume average of $\dot\delta$ over one hole (with
$\delta\equiv\rho/\bar{\rho}-1$) evolves more slowly.

The shell in the open case and the centre in the closed case are
in the non-linear regime, as shown in figure \ref{fig:expansion}, so the
Rees--Sciama effect \cite{Rees:1968} is expected to be
important in addition to the linear ISW effect.
The spherical symmetry leads to stronger cancellation than
in ray-tracing through simulated structures \cite{Seljak:1995,Cai:2010},
where the typical amplitude of the Rees--Sciama effect is $\sim10^{-6}$,
one order of magnitude above our open model and two or
three orders of magnitude above our closed model.
In the case of a central underdensity, the linear ISW effect
and the Rees--Sciama effect enhance each other, whereas for
a central overdensity, they pull in opposite directions
\cite{Cai:2010}, which may partly explain why the temperature
signal in the closed case is smaller than in the open case.

However, the difference in amplitude also reflects the fact that the
maximum value of $|E|$ is smaller in the closed model, to avoid
shell crossings. To better understand the differences in the two
models, let us consider the proper volume averages
$\langle\rangle_V$ of $\Delta H/\bar{H}\equiv H/\bar{H}-1$
and $\dot{\delta}=-3(1+\delta)\Delta H$ over a single hole,
given in table \ref{table:params}.
Both models expand on average a little slower than the background
at present day, $\langle\Delta H/\bar{H}\rangle_V=-2.0 \cdot 10^{-7}$
in the open case and $-1.8 \cdot10^{-9}$ in the closed case.
These numbers are not far from the maximum temperature amplitudes
(though note that $\Delta T/\bar{T}$ can have different sign in the
open and closed models).
The redshift can be expected to be calculable from the average
expansion rate when passing through many structures,
but not necessarily in the case of one structure, and it is also
important to take the dust shear into account
\cite{Rasanen:2008be, Rasanen:2009, Rasanen:2011b, Lavinto:2013},
so such a close agreement can be fortuitous, but the order of magnitude
is right. 
In the closed model, the local expansion rate is always positive, whereas
in the open model, the volume element in the overdense shell shrinks at late
times, as shown in figure \ref{fig:expansion}.

In linear perturbation theory, the temperature perturbation due to
the ISW effect is
\be
  \frac{\Delta T}{\bar{T}} = 2 \int \D t \dot{\Phi} \ ,
\ee

\noindent where the integral is taken along the null geodesic,
and the metric perturbation $\Phi$ is proportional to $\delta$ with a
growth function that varies slowly compared to the time it takes
to cross a hole. As long as null geodesics sample the
spacetime volume fairly, we expect the difference in
$\langle\dot{\delta}\rangle_V$
to be indicative of the difference in the temperature
perturbation (as noted above, it is a priori not obvious that
such an argument applies in the case of a single hole).
We have $\langle\dot{\delta}\rangle_V=0.156 H_0$ in the open model
and $0.010 H_0$ in the closed model, a ratio of 16, in rough
agreement with the ratio of temperature anisotropies.

Let us mention that nonlinear voids have been considered as a source
of large temperature anisotropies in the CMB
\cite{MG:1990a, MG:1990b, Panek:1992, Arnau:1993, Fullana:1996, TomitaInoue:2008},
in particular as a way to explain the so-called Cold Spot. However,
no sufficiently large and deep void has been detected, and its existence
would be extremely unlikely in the $\Lambda$CDM model
\cite{InoueSilk:2006, Rudnick:2007, Cruz:2008, Sakai:2008, Masina:2008, Smith:2008, Masina:2009, Granett:2009, Bremer:2010, Cai:2010, Inoue:2010, Inoue:2011, Szapudi:2014, Finelli:2014, Nadathur:2014}.
It has also been argued that the impact of large underdense and overdense
regions on the CMB is significantly stronger than predicted in linear theory
in the $\Lambda$CDM model \cite{Granett:2008a, Granett:2008b, Flender:2012},
though subsequent studies have found a signal in agreement with such
predictions \cite{Cai:2013, Hotchkiss:2014, Granett:2015}.
For constraints on exceptional voids from the CMB power spectrum,
see \cite{Zibin:2014}.

\begin{figure}
\includegraphics[width=0.49\linewidth]{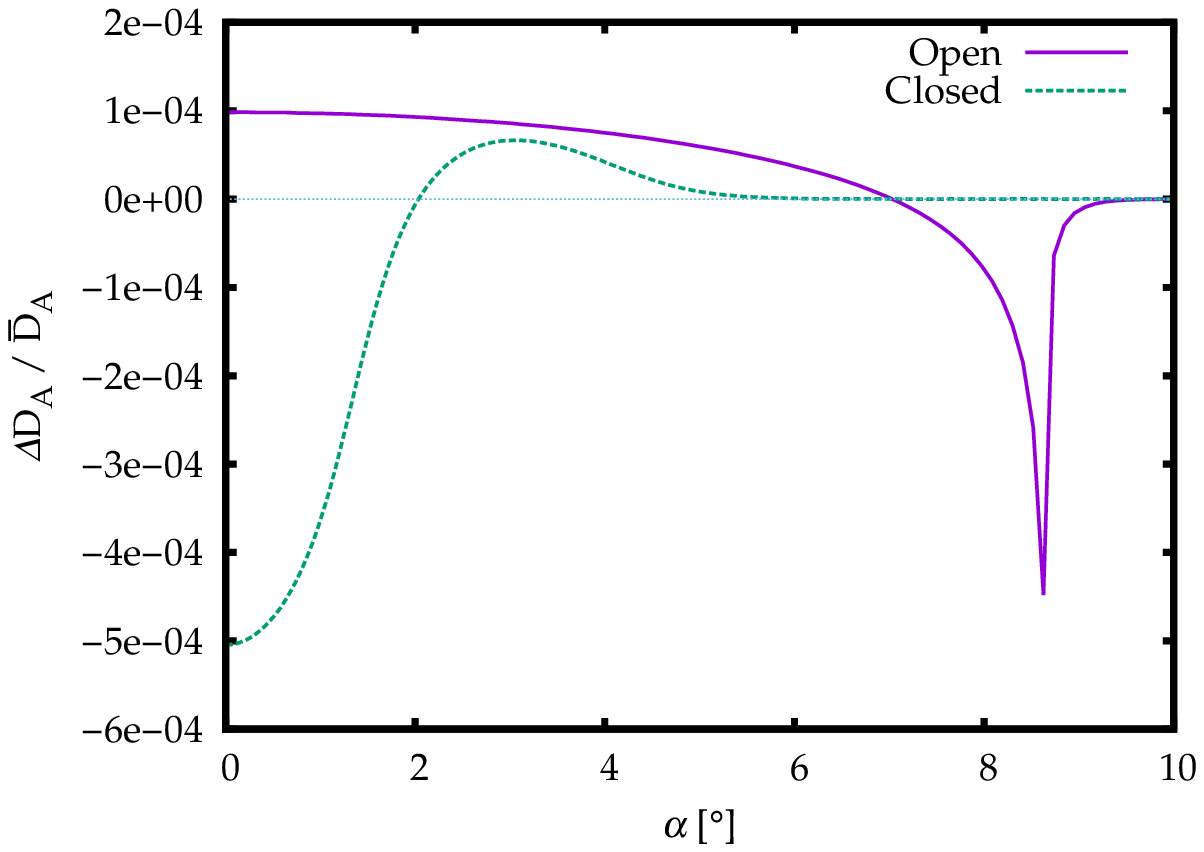} 
\includegraphics[width=0.49\linewidth]{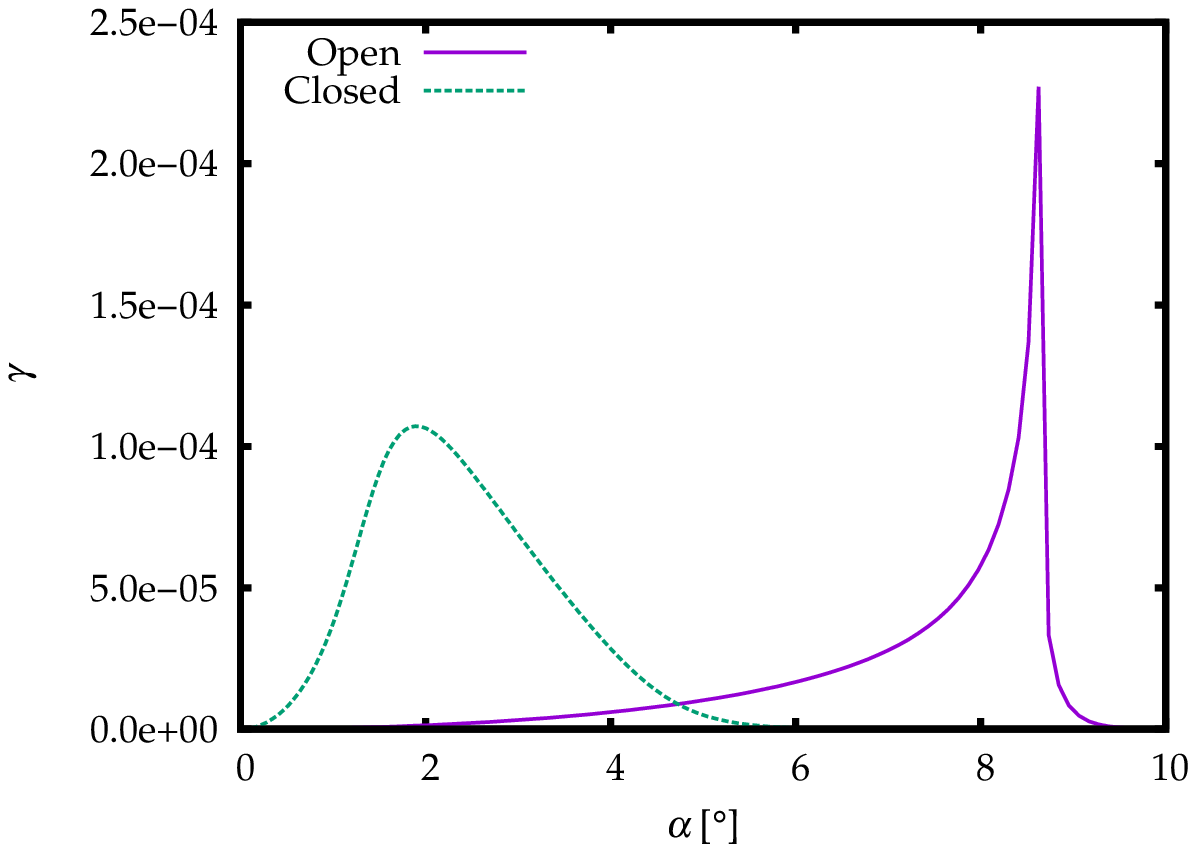}
\caption{The angular profiles of $\Delta D_A/\bar{D}_A$ (left) and $\gamma$
(right) for the open and closed models as a function of the viewing angle
$\alpha$ for a single hole, with its centre located at a coordinate distance
of $200 \mpc$.}
\label{fig:Dprofiles}
\end{figure}

\begin{figure}
\includegraphics[width=0.5\linewidth]{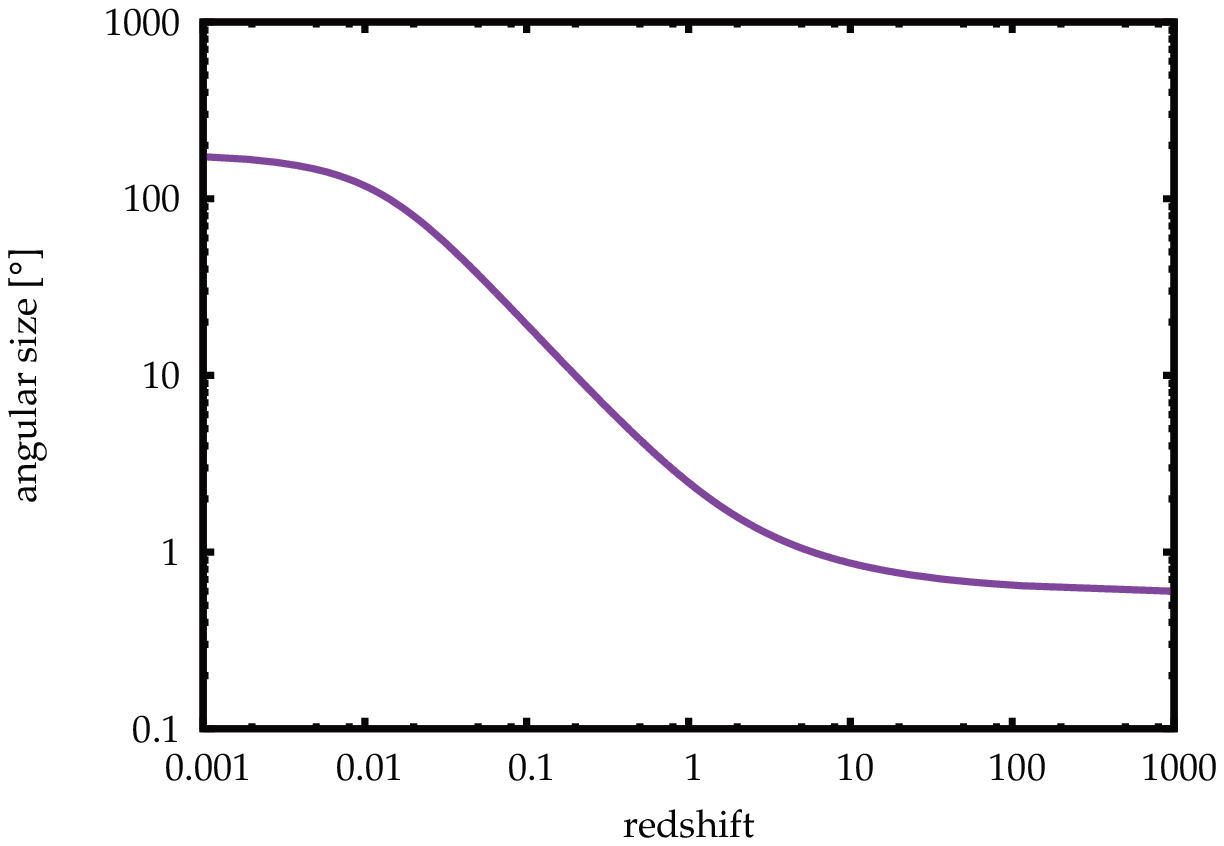}
\includegraphics[width=0.49\linewidth]{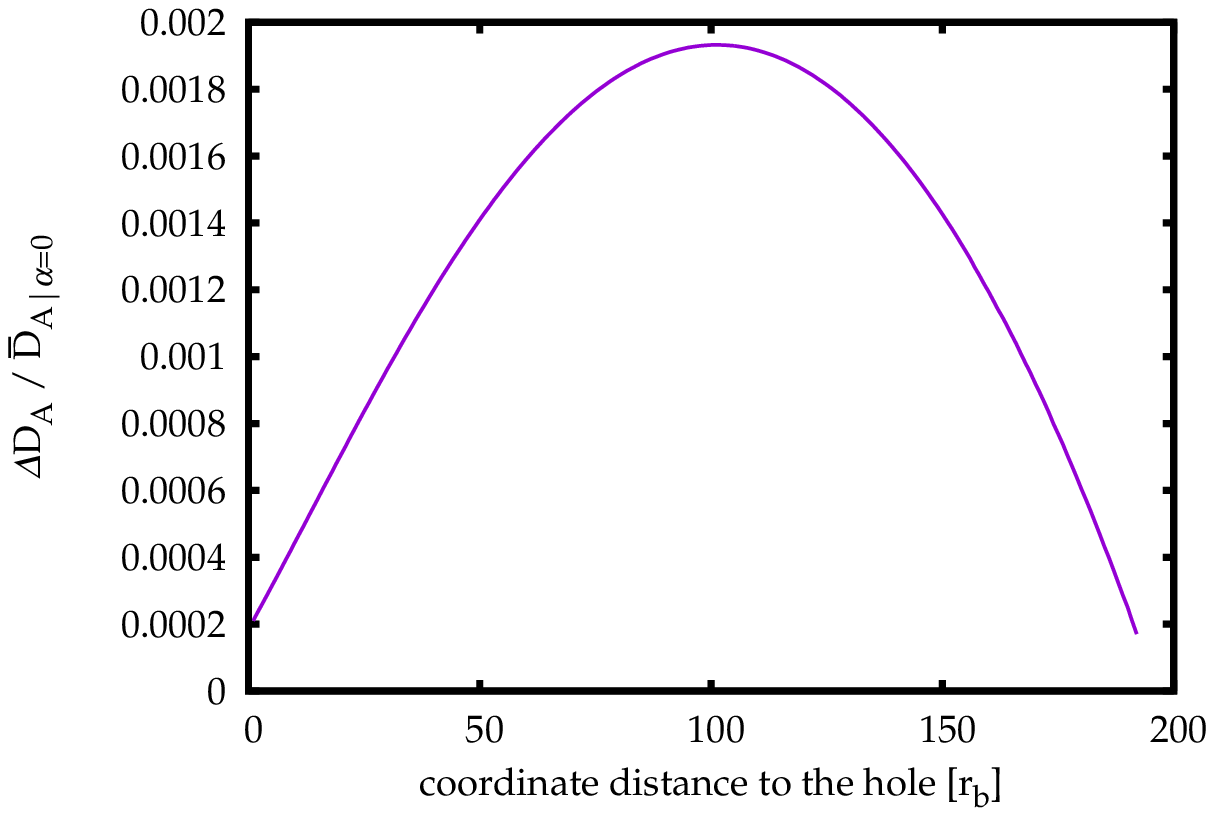}
\caption{The angular size (left) of a hole ($r_b = 50 \mpc$) as a function of redshift to the hole centre. The inhomogeneities have radii $\sim 30 \mpc$ and $\sim 20 \mpc$ in the open and closed models, respectively, so they are somewhat smaller. Temperature shift (right) for a radial light ray as a function of the distance to the centre in units of $r_b$, for the open model.}
\label{fig:distance}
\end{figure}

In the open model, the distance perturbation is positive in the centre
with a distinct negative spike corresponding to the overdense shell,
as shown in figure \ref{fig:Dprofiles}.
That is, the image is demagnified in the centre and magnified on the
shell. The situation is reversed in the closed model, though
the profile is smoother, as in the case of the temperature.
The maximum amplitude varies from $\sim10^{-4}$ to $\sim10^{-3}$,
depending on the distance to the hole.
In contrast to temperature perturbations, which are mostly generated by
holes close to the observer, perturbations in the angular diameter
distance are most strongly produced by holes halfway between source
and observer, a common feature in gravitational lensing
(see e.g. \cite{P.Schneider1992}, p. 25). 
The variation with distance is shown in figure \ref{fig:distance}
for an open hole; the closed case behaviour is similar.
For discussion of lensing by voids, see
\cite{Bolejko:2012, Mood:2013, Chen:2013}.

The amplitude of the distance perturbation is less sensitive to the choice
of profile than the amplitude of the temperature perturbation.
The open and closed model results are comparable in amplitude, but have
distinct dependence on the viewing angle. There is an argument that the
maximum amplitude of the deviation would be of the order $(H_0 r_b)^3$
for the temperature and $(H_0 r_b)^2$ for the distance
\cite{Sugiura:1999b, Brouzakis:2008}.
Our maximum amplitude of $10^{-6}$ for the temperature perturbation agrees with
this scaling for nearby open holes (for closed holes, the amplitude
is smaller). The distance perturbation is indeed $10^{-4}\sim(H_0 r_b)^2$
for holes far from the midpoint between observer and the last
scattering surface, but for holes at the optimal distance, it is
an order of magnitude larger.

The amplitude of the integrated null shear in both the open
and closed cases is $\gamma\sim10^{-4}$.
As with the distance, the angular profile is smoother in the closed case,
with the open case characterised by a spike corresponding to the light ray
passing through the overdense shell, as shown in figure \ref{fig:Dprofiles}.

\begin{table}\begin{center}
\begin{tabular}{c|c c}
 & Open & Closed \\ 
\hline 
$\langle \Delta T / \bar{T} \rangle$ & $ (2.15 \pm 0.10) \times 10^{-8} $  & $(3.17 \pm 0.03) \times 10^{-9} $\\
$\sigma_{\Delta T/\bar{T}}$ & $ (9.05 \pm 0.10) \times 10^{-8}$ & $(2.13 \pm 0.04)\times 10^{-9} $ \\
$\langle \Delta D_A / \bar{D}_A \rangle$ & $(8.5 \pm 4.3) \times 10^{-5} $ & $(1.1 \pm 2.4) \times 10^{-5} $ \\ 
$\langle (\Delta D_A / \bar{D}_A)^2 \rangle$ & $(1.65 \pm 0.02) \times 10^{-5} $ & $(5.28 \pm 0.08) \times 10^{-6} $ \\ 
$\sigma_{\Delta D_A/\bar{D}_A}$ & $(4.07 \pm 0.03) \times 10^{-3} $ & $(2.30 \pm 0.02) \times 10^{-3} $ \\
$\langle \gamma \rangle$ & $(1.78 \pm 0.01) \times 10^{-3}$ & $(9.85\pm 0.08) \times 10^{-4}$ \\
$\sigma_{\gamma}$ & $(1.01 \pm 0.02)\times 10^{-3}$ & $(6.03 \pm 0.11)\times 10^{-4}$ \\
$\langle\mu\rangle - 1$ & $(-1.20 \pm 0.85) \times 10^{-4}$ & $(-0.5 \pm 4.7) \times 10^{-5}$ \\
$\langle\mu^{-1}\rangle - 1$ & $(1.86 \pm 0.85  ) \times 10^{-4}$ & $( 2.7\pm 4.8) \times 10^{-5}$
\end{tabular} 
\caption{Mean shift and variation for a single beam for the temperature, distance and shear, as well as the average of the square of the distance deviation, magnification and inverse magnification. Errors are calculated using a bootstrap algorithm, and give one standard deviation. The probability distribution for $\langle \Delta D_A / \bar{D}_A \rangle$ is plotted in figure \ref{fig:bootstrap}.}
\label{table:errors}
\end{center}
\end{table}

\paragraph{Multiple holes.}

As the calculation is too time-intensive to do on very small
scales, we focus on multipoles $l\lesssim100$, corresponding to
angles $\gtrsim1^{\circ}$. The angular size of a single hole
is plotted in figure \ref{fig:distance}. Therefore we
do not have to populate the universe all the way to the
last scattering surface, we only need to include holes up to redshift
$z\sim6$, corresponding to an emission time of $\sim1$ Gyr and a
distance of $\sim 6000\mpc$.
This is beyond half-way to the last scattering surface
(the distance to $z=1090$ is 9600$\mpc$), so we would not expect to
find considerably more power on small scales even if we had holes
all the way up to $z=1090$.

\begin{figure}
\includegraphics[width=0.325\textwidth]{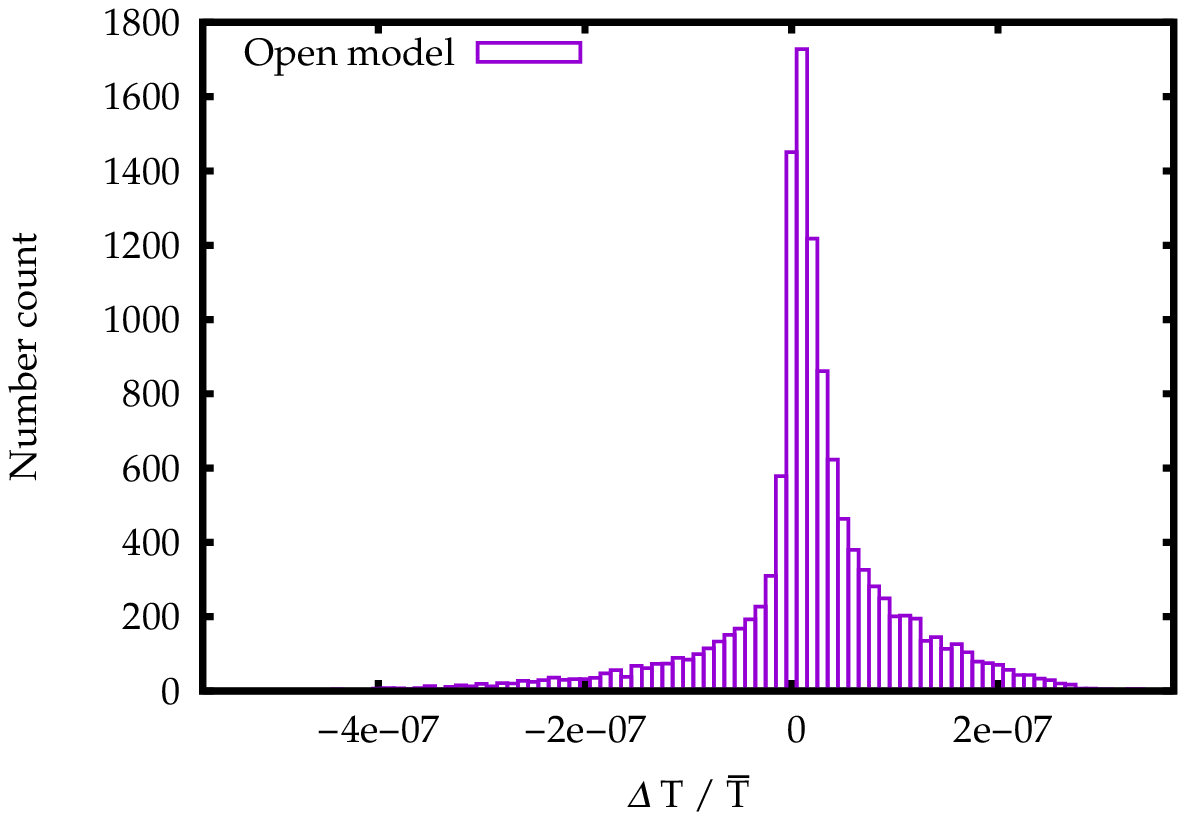}
\includegraphics[width=0.325\textwidth]{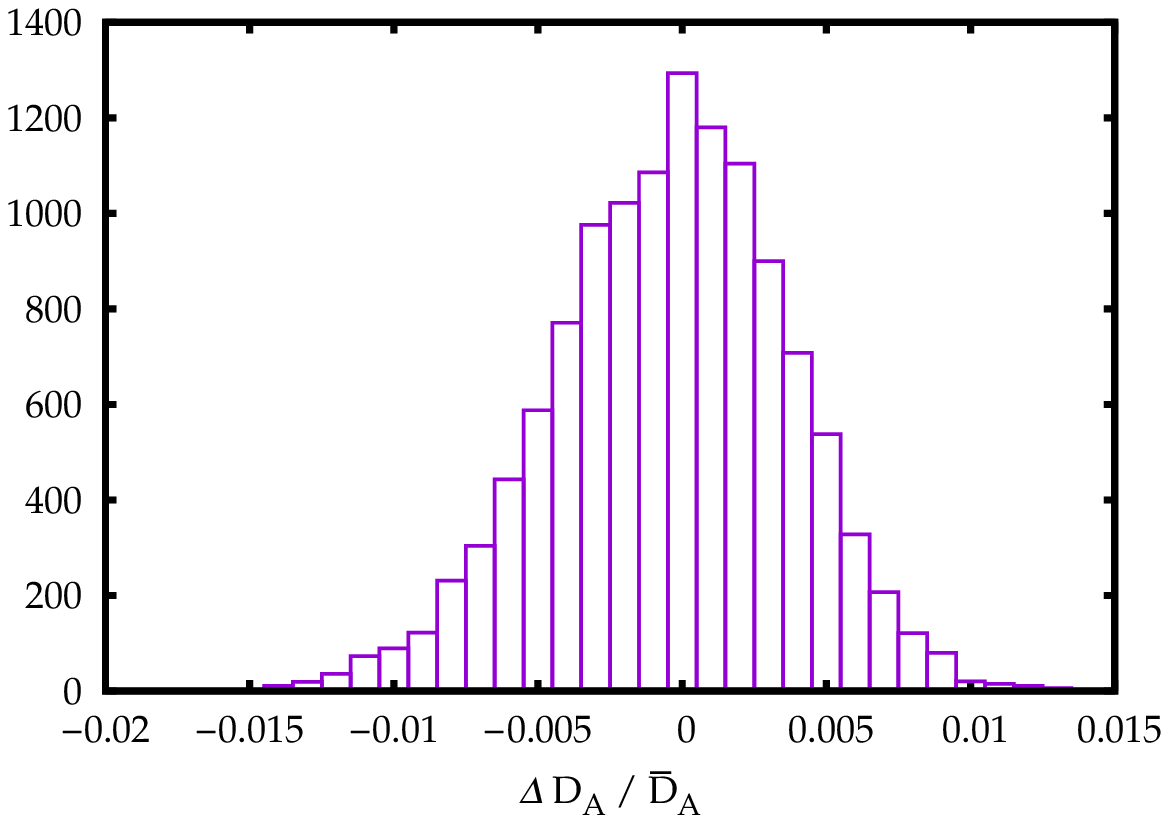}
\includegraphics[width=0.325\textwidth]{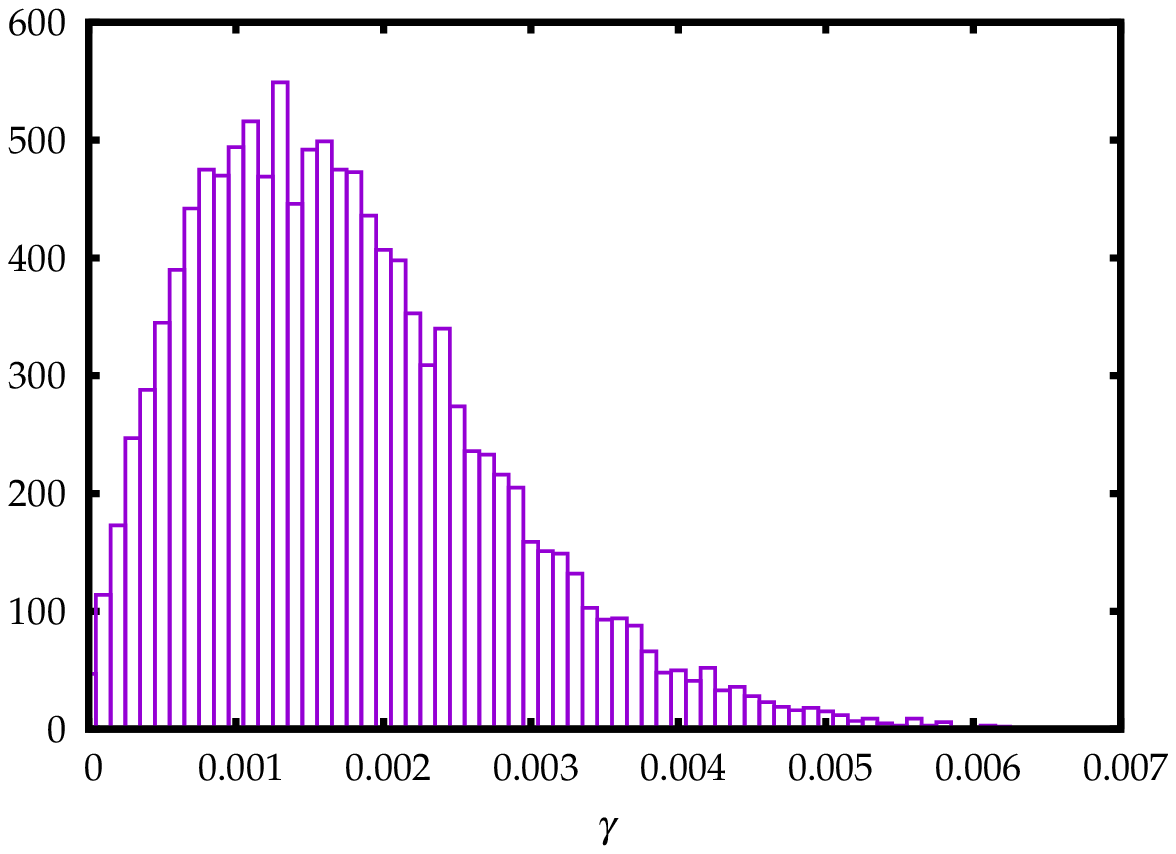} \\ 
\includegraphics[width=0.325\textwidth]{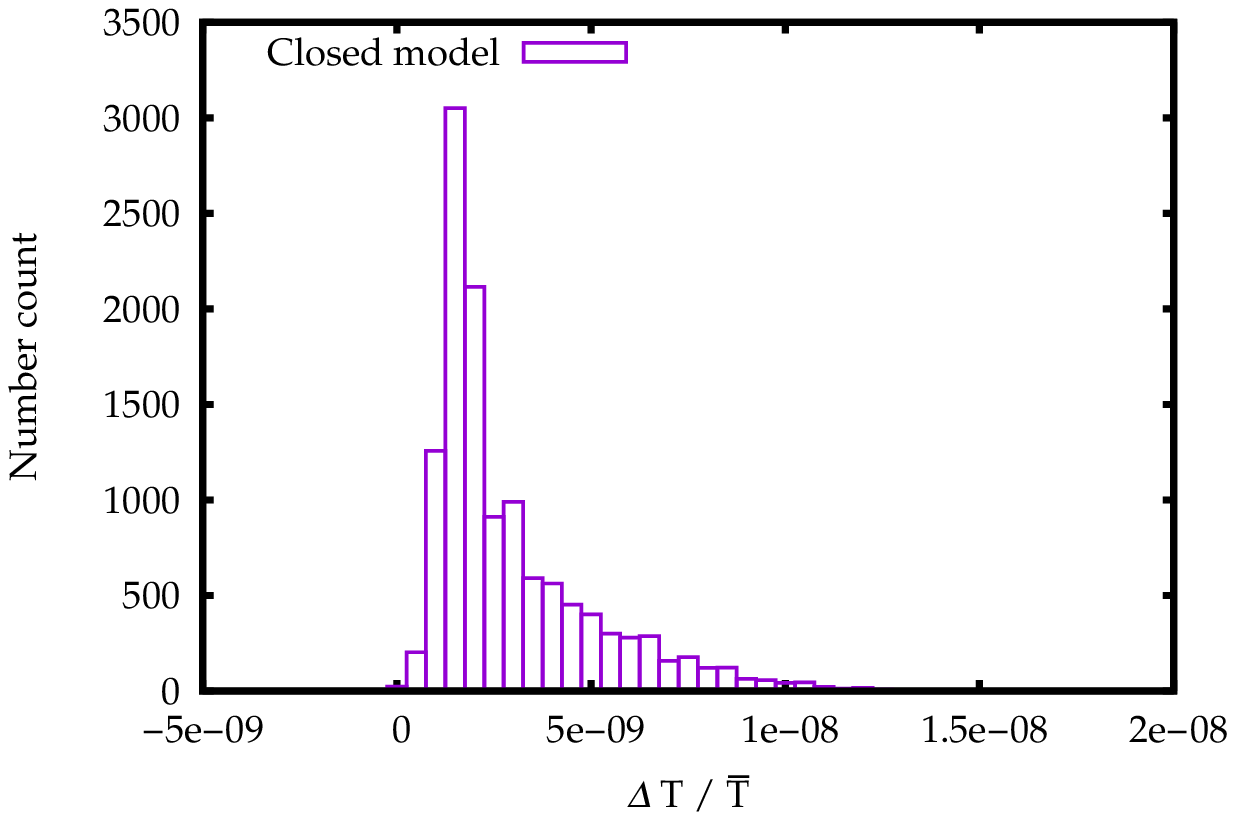}  
\includegraphics[width=0.325\textwidth]{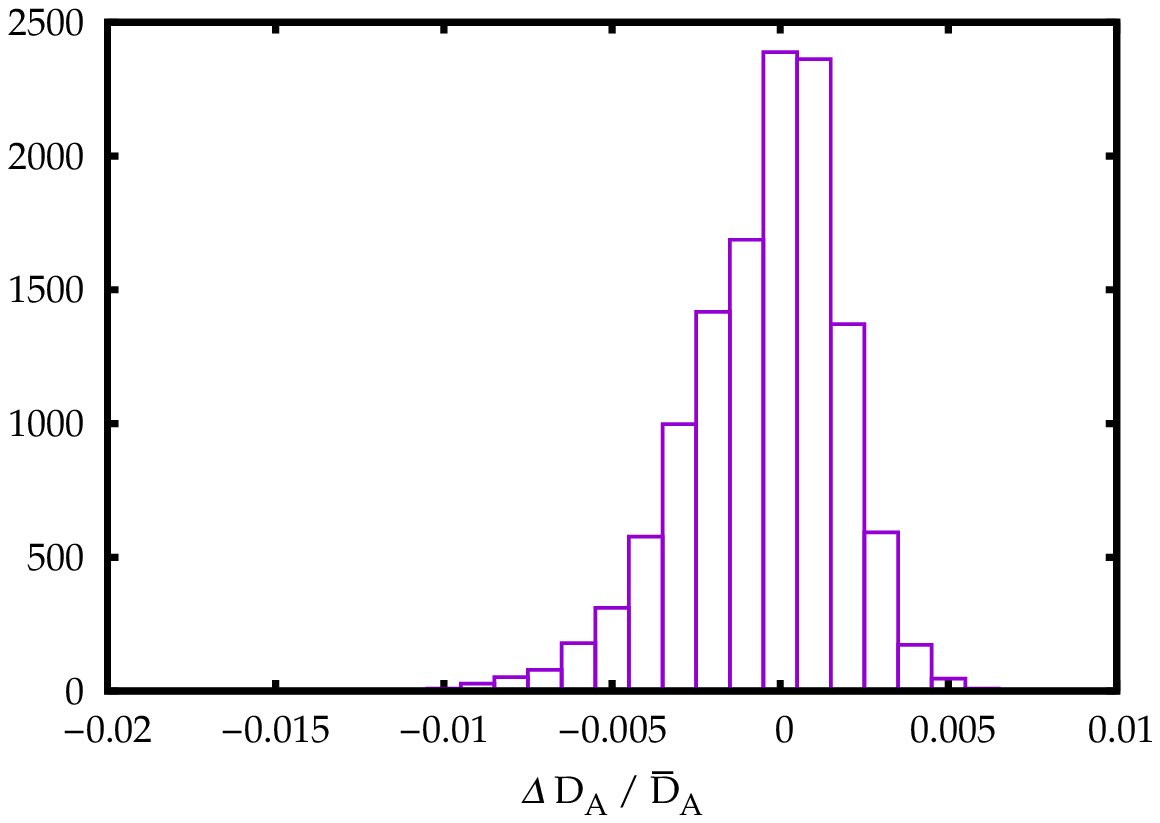} 
\includegraphics[width=0.325\textwidth]{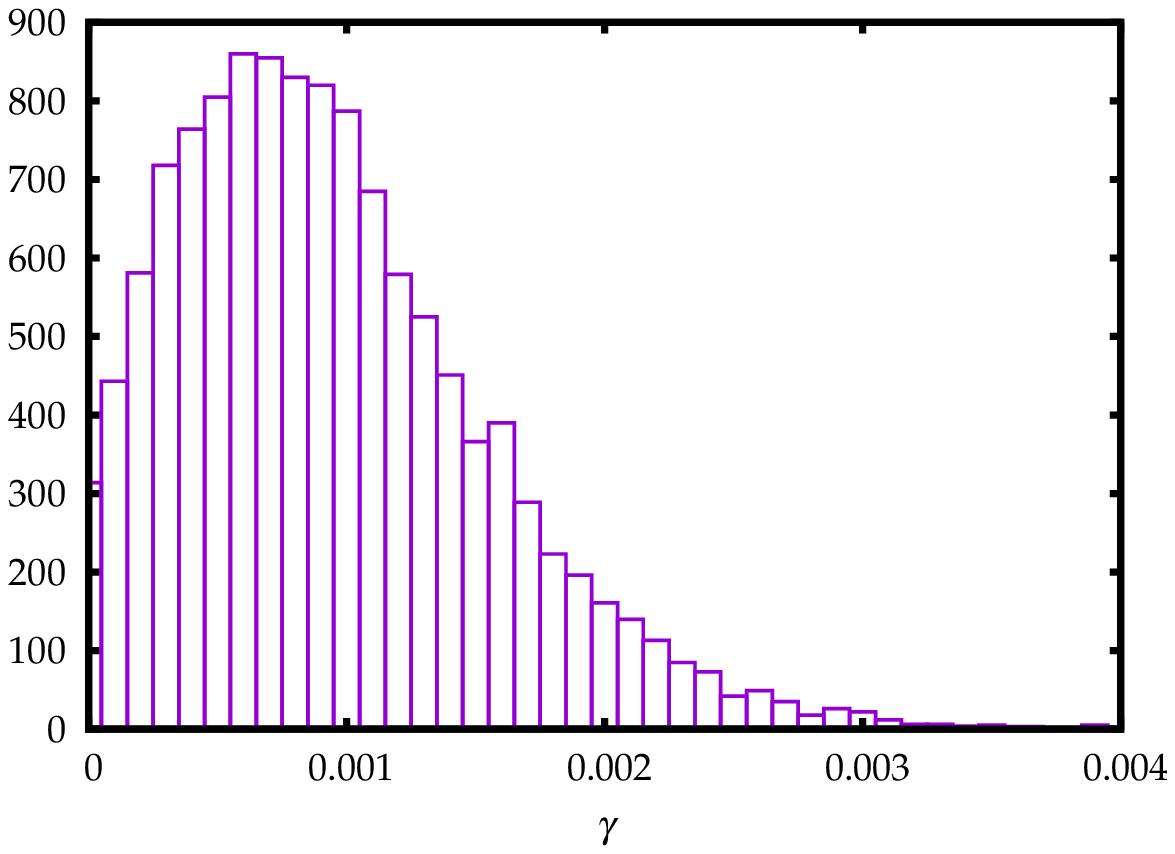}
\caption{Histograms of $\Delta T/\bar{T}$ (left), $\Delta D_A/ \bar{D}_A$ (middle) and $\gamma$ (right) for open (top) and closed (bottom) models. The mean values and standard deviations are given in table \ref{table:errors}.}
\label{fig:histograms}
\end{figure}

\begin{figure}
\center{\includegraphics[width=0.5\textwidth]{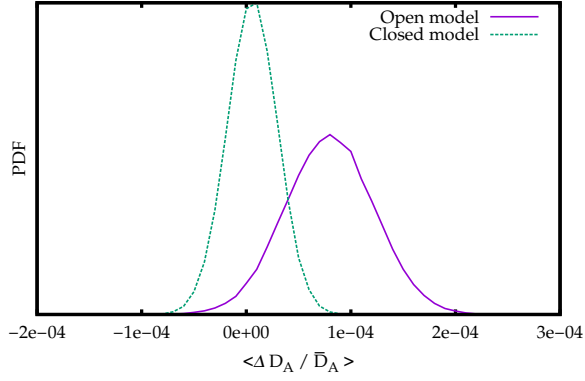}}
\caption{Probability density function for $\langle\Delta D_A / \bar{D}_A\rangle$, estimated using a bootstrap algorithm.}
\label{fig:bootstrap}
\end{figure}

We propagate $N=$ 12 288 beams.
In table \ref{table:errors} we report the values
of $\Delta T/\bar{T}$, $\Delta D_A/\bar{D}_A$ and $\gamma$
averaged over the sky, with errors calculated using a bootstrap
algorithm where random sky maps are generated from the original
simulated map.
We also give the variation for a single beam for the same quantities.
We also give the averages of $(\Delta D_A/\bar{D}_A)^2$, $\mu$ and $\mu^{-1}$.
If the sky pixels were independent, we would expect the
error on the mean to be roughly $1/\sqrt{N}$ times the
variation for a single beam, and the numbers in table \ref{table:errors}
agree quite well with this.
The histograms showing the detailed shape of the
distribution for single beams are plotted in figure \ref{fig:histograms},
and the bootstrap errors for $\langle\Delta D_A/\bar{D}_A\rangle$
are shown in \ref{fig:bootstrap}.

The mean temperature shift is
$\langle\Delta T/\bar{T}\rangle=(2.15\pm0.10)\times10^{-8}$
in the open case and $(3.17\pm0.03)\times10^{-9}$,
an order of magnitude smaller, in the closed case.
(Our errors always correspond to one standard deviation.)
For the open model, the hot and cold regions cancel to high 
precision, so the mean value is well below the typical
fluctuation range $\sim10^{-7}$ of a single beam.
For the closed model, the temperature perturbation is 
everywhere positive, so there are no cancellations and the 
mean is of the same order of magnitude as a typical fluctuation. 
As the effective packing fraction is small and the temperature
signal falls off as a function of distance to the hole,
as shown in figure \ref{fig:Tdistance}, the temperature perturbation
along a single line of sight typically gets contributions from only
a few holes (more in the closed than in the open case).
Therefore the typical fluctuation is much smaller than the maximum
fluctuation for a single hole (obtained for a radial light ray), in both cases.

The mean shift of the angular diameter distance is 
$\langle\Delta D_A/\bar{D}_A\rangle=(8.5\pm4.3)\cdot10^{-5}$
in the open model and $(1.1\pm2.4)\cdot10^{-5}$ in the closed model.
In contrast to the temperature, the mean shift is not statistically
distinguishable from zero, so we can only quote the 95\% upper limit
$|\langle\Delta D_A/\bar{D}_A\rangle|\lesssim10^{-4}$, of the same
order of magnitude as the signal for a single hole.
In both cases, the variation for single beams is at the level
$10^{-3}$, which explains the large errors.
Because of this large variation, it is particularly important
to evaluate the errors in the case of the distance, and not just
consider the shift in the mean evaluated over a single sky.
If the shifts $\langle\Delta D_A/\bar{D}_A\rangle$ were the mean
values reported above, we would expect to see them clearly with
3 times more beams in the open case and 50 times more in the closed case.

The mean integrated null shear is $\langle\gamma\rangle\approx10^{-3}$
in both the open and closed case, of the same order as the typical
result for a single ray. The reason is that $\gamma$ is by definition
non-negative, so there can be no cancellations across the sky.
The sky average $\langle\gamma\rangle$ is also about an order of
magnitude larger than the result for a single hole.
The variation for single beams is of the same order of magnitude,
so with 12 288 beams, the error on the mean is below 1\%.

\begin{figure}
\includegraphics[width=0.49\linewidth]{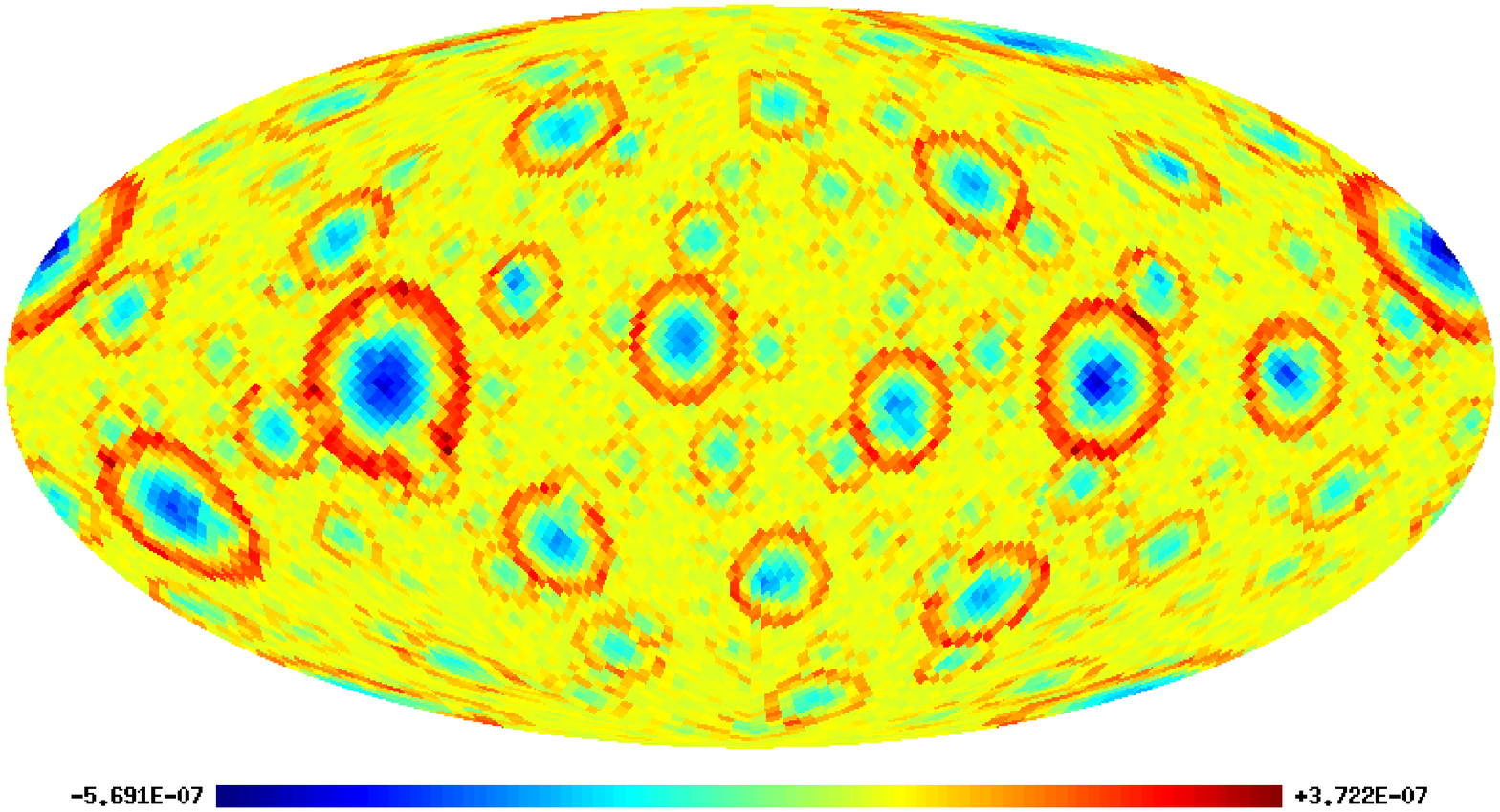} 
\includegraphics[width=0.49\linewidth]{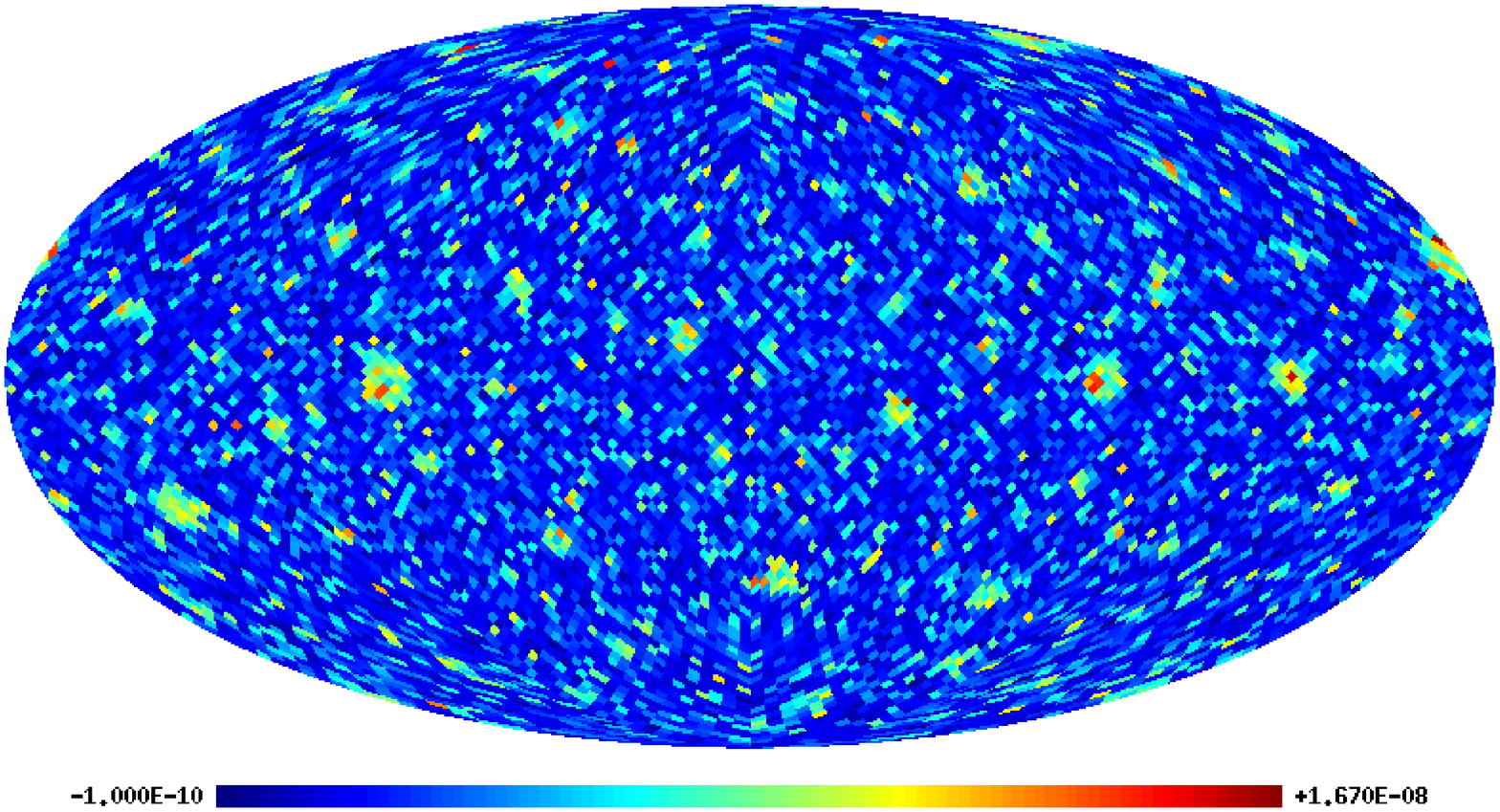} \\
\includegraphics[width=0.49\linewidth]{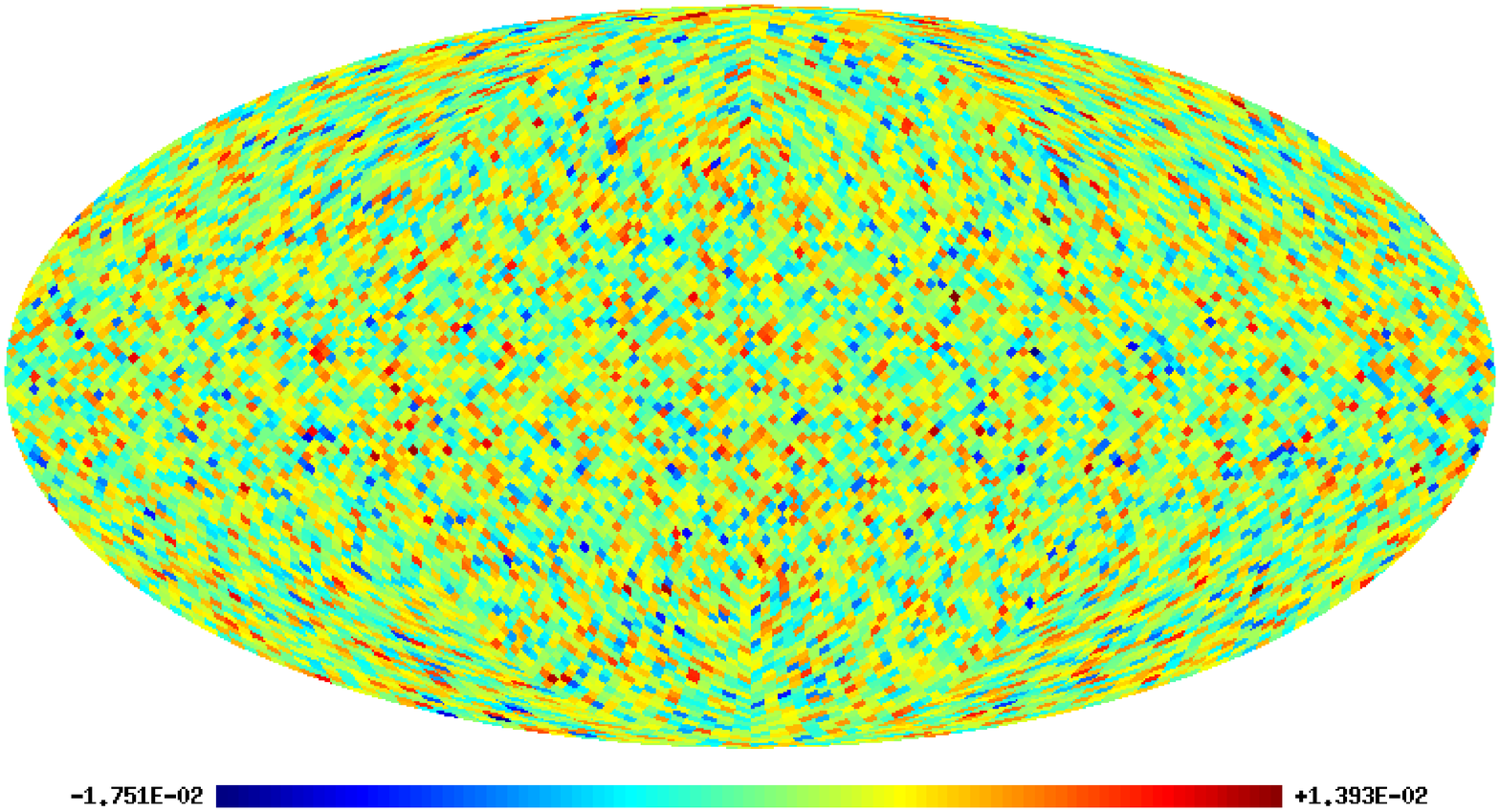} 
\includegraphics[width=0.49\linewidth]{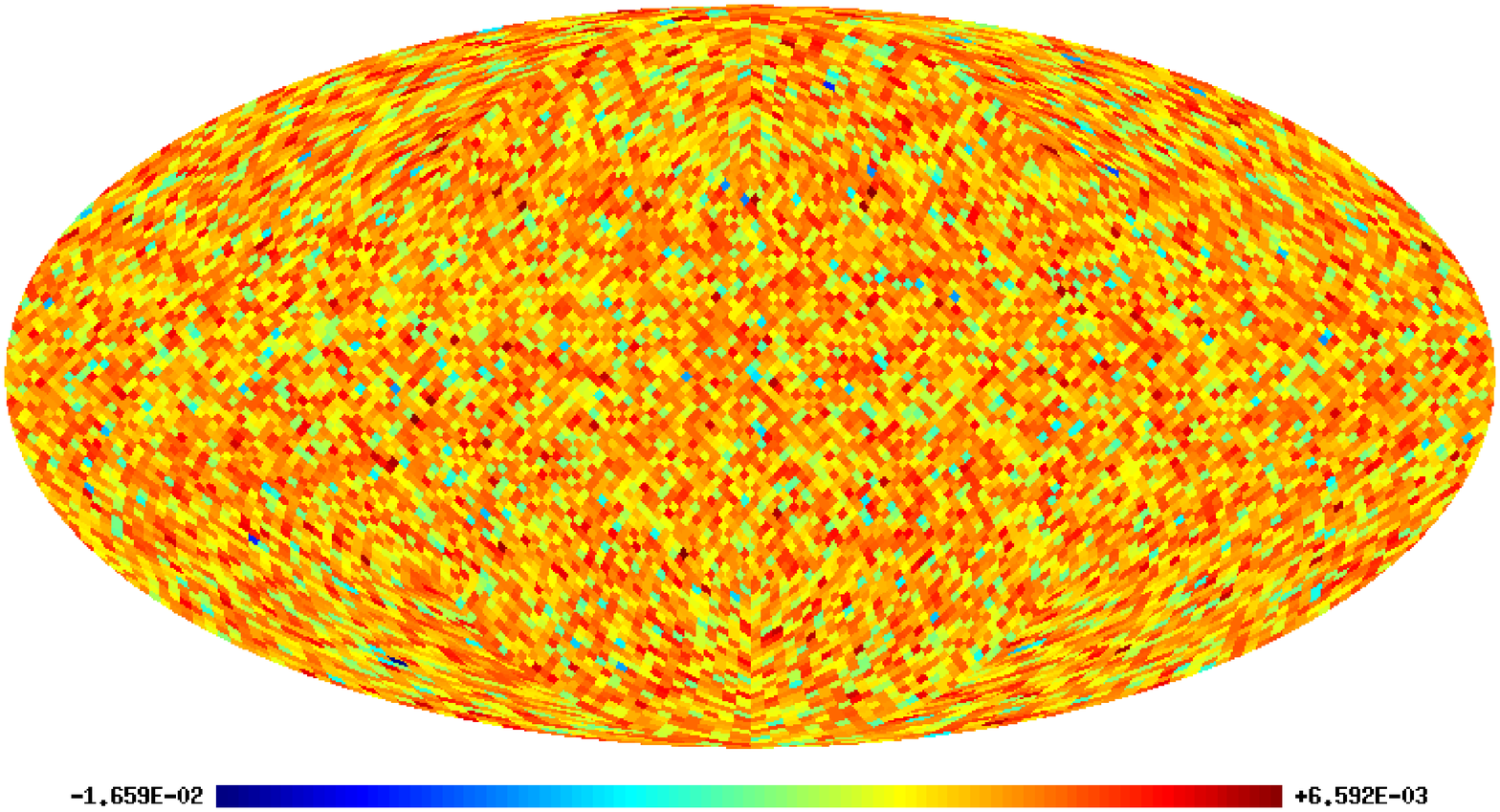} \\
\includegraphics[width=0.49\linewidth]{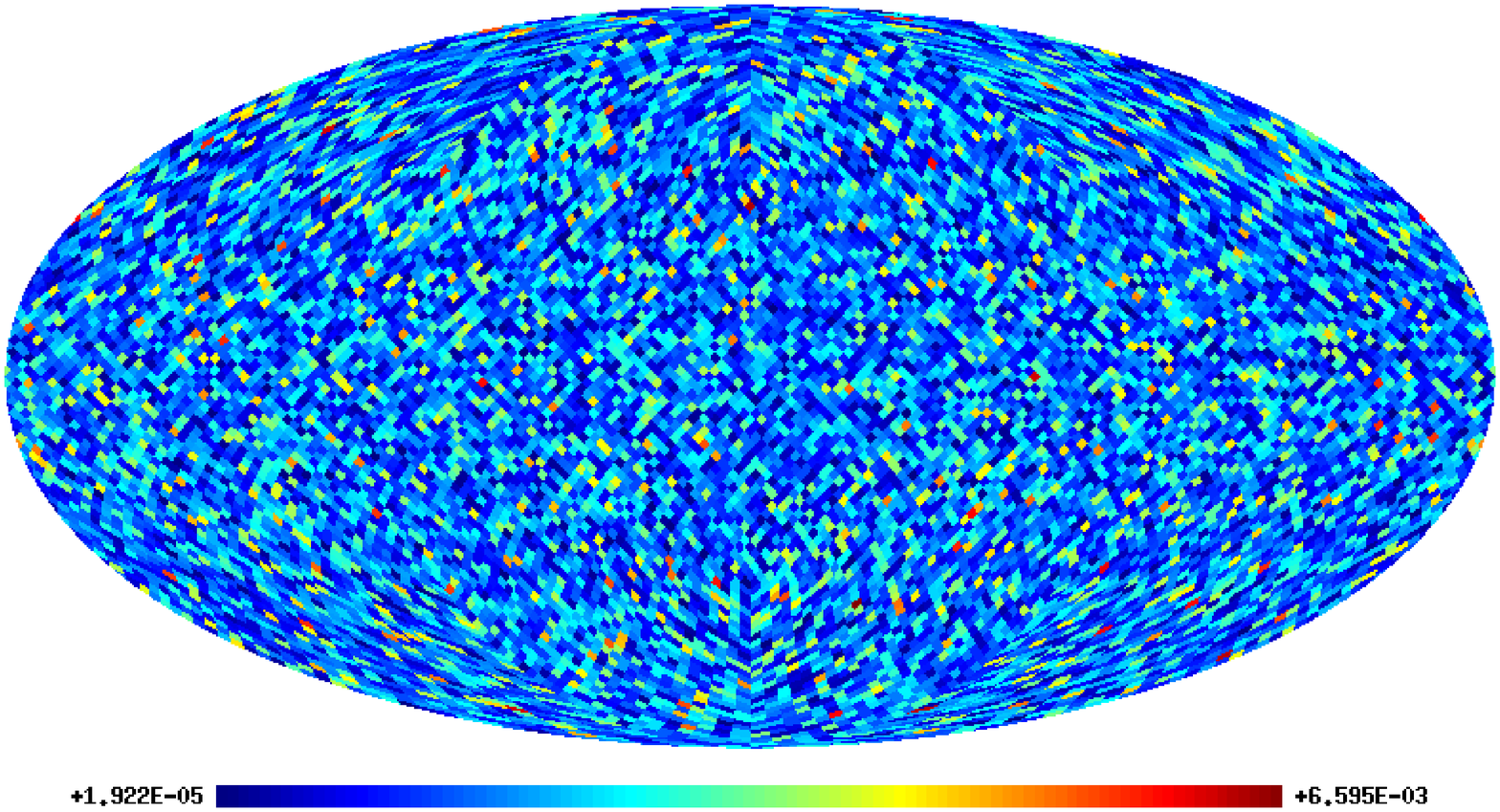} 
\includegraphics[width=0.49\linewidth]{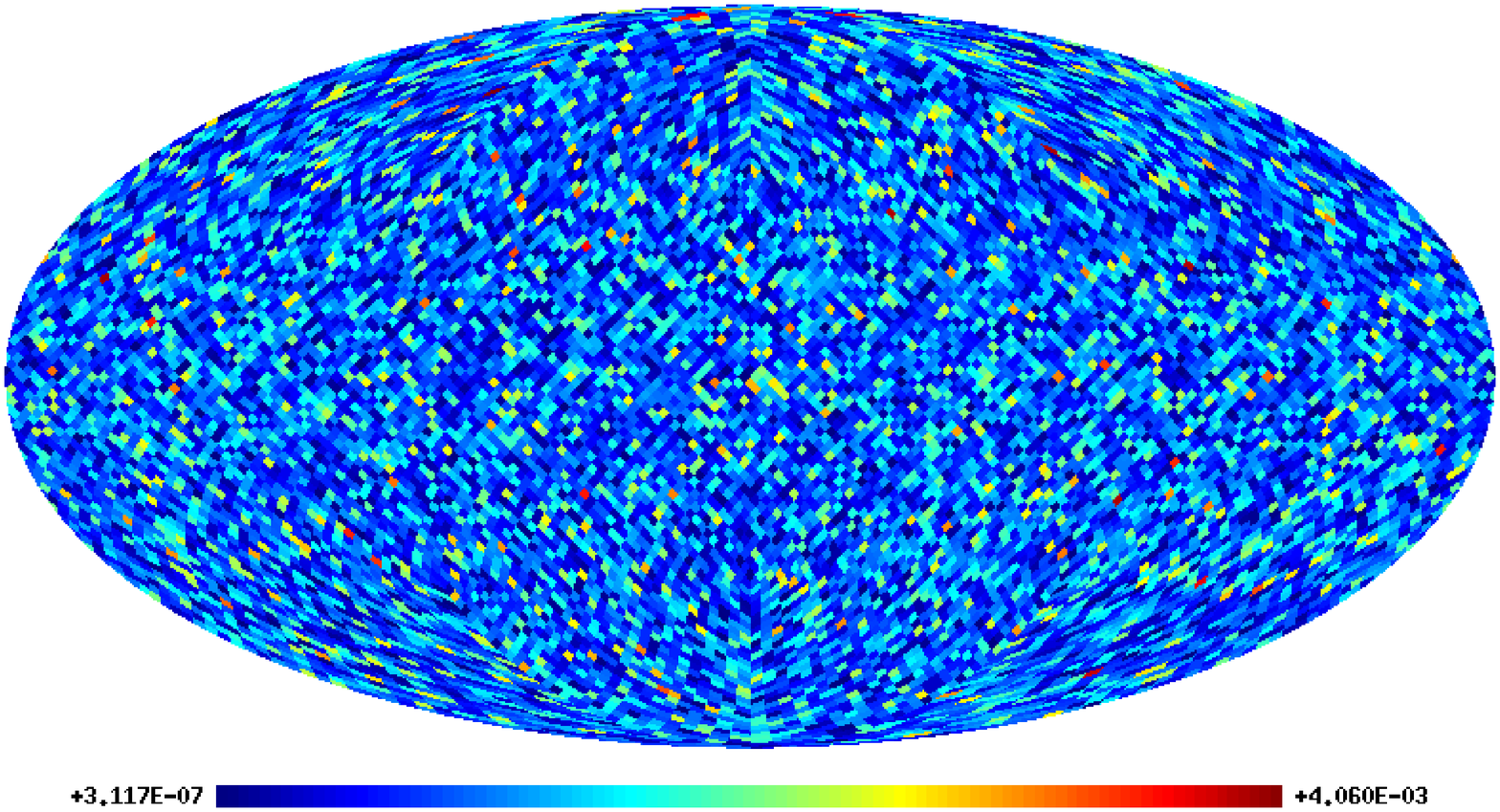} 
\caption{Maps of $\Delta T/\bar{T}$ (top), $\Delta D_A/ \bar{D}_A$ (middle) and $\gamma$ (bottom) for the open (left) and closed (right) model.}
\label{fig:maps}
\end{figure}

We calculate power spectra from the CMB maps using the HEALPix
\cite{Gorski:2004by} package. HEALPix splits the sky into $12$ equal
sized regions, all of which have $N^2$ pixels. We used $N=32$, giving
$N_{\mathrm{pix}}=$ 12 288 pixels in total, so we have one beam per pixel.
This is enough to calculate the power spectra up to $l < 3N$, though
the statistical errors are considerable for $l > 2N$.
The resulting sky maps for $\Delta T/\bar{T}$, $\Delta D_A/\bar{D}_A$
and $\gamma$ are shown in figure \ref{fig:maps}.
As the holes are spherically symmetric, the maps are missing dipolar
structures characteristic of the Rees--Sciama effect \cite{Cai:2010};
it would be possible to model them by using non-spherically symmetric
Szekeres holes.
Already by eye it is clear that the temperature perturbations
are concentrated on large scales, whereas distance
and shear have most of their power on small scales.
We quantify this by computing the power spectrum for
$f(\theta, \phi)=\Delta T/\bar{T}, \Delta D_A/\bar{D}_A, \gamma$
by expanding in terms of the spherical harmonics $Y_{lm}$,
\begin{align}
C_l^{ff} &\equiv \frac{1}{2l+1} \sum_{m=-l}^{l} |a_{lm}^f|^2 \\
a_{lm}^f &\equiv \int_0^{\pi} \D\theta \int_0^{2\pi} \D\phi \ Y_{lm}^*(\theta,\phi) f(\theta,\phi) \ .
\end{align}
The coefficients $a_{lm}$ are evaluated from the maps by summing over the pixels $p$,
\be a_{lm} = \frac{4\pi}{N_{\mathrm{pix}}} \sum_{p=0}^{N_{\mathrm{pix}}-1} Y_{lm}^*(\theta_p, \phi_p) f(\theta_p, \phi_p) \ .
\ee

\begin{figure}
\begin{center}
\includegraphics[width=0.5\linewidth]{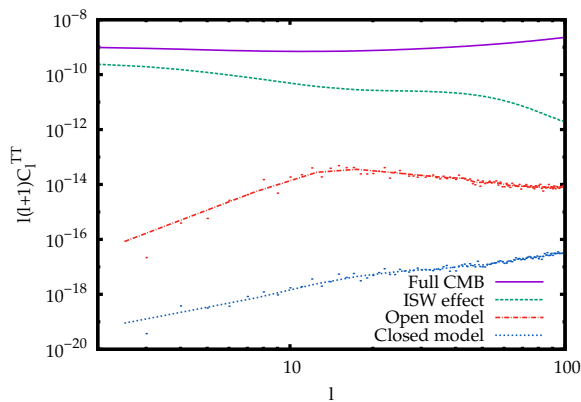}  
\end{center}
\caption{Power spectrum of the temperature anisotropy $\Delta T/\bar{T}$. Plotted are the usual $\Lambda$CDM power spectrum, the late-time linear ISW contribution to the $\Lambda$CDM result and the power spectra from our Swiss Cheese models. The lines are binned spectra and the dots are individual multipoles.}
\label{fig:cltt}
\end{figure}

\begin{figure}
\includegraphics[width=0.49\linewidth]{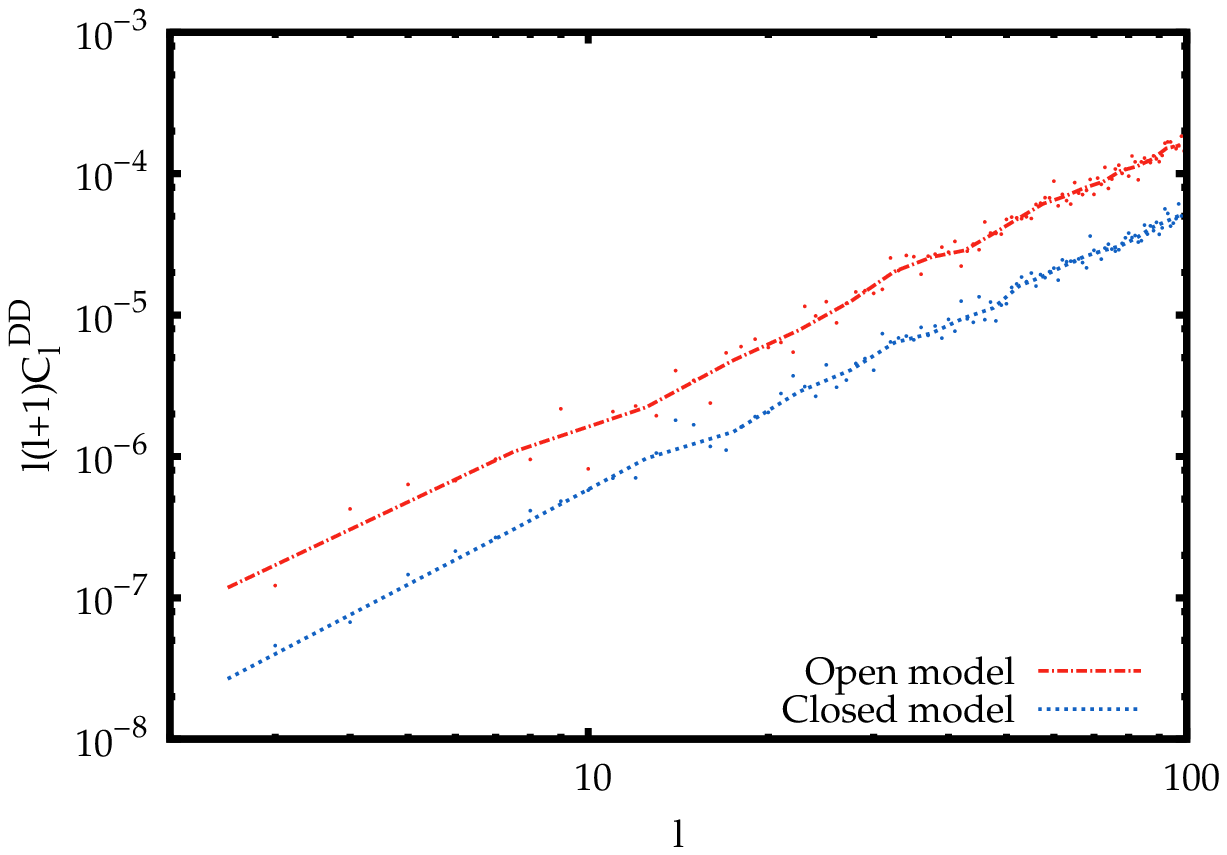} 
\includegraphics[width=0.49\linewidth]{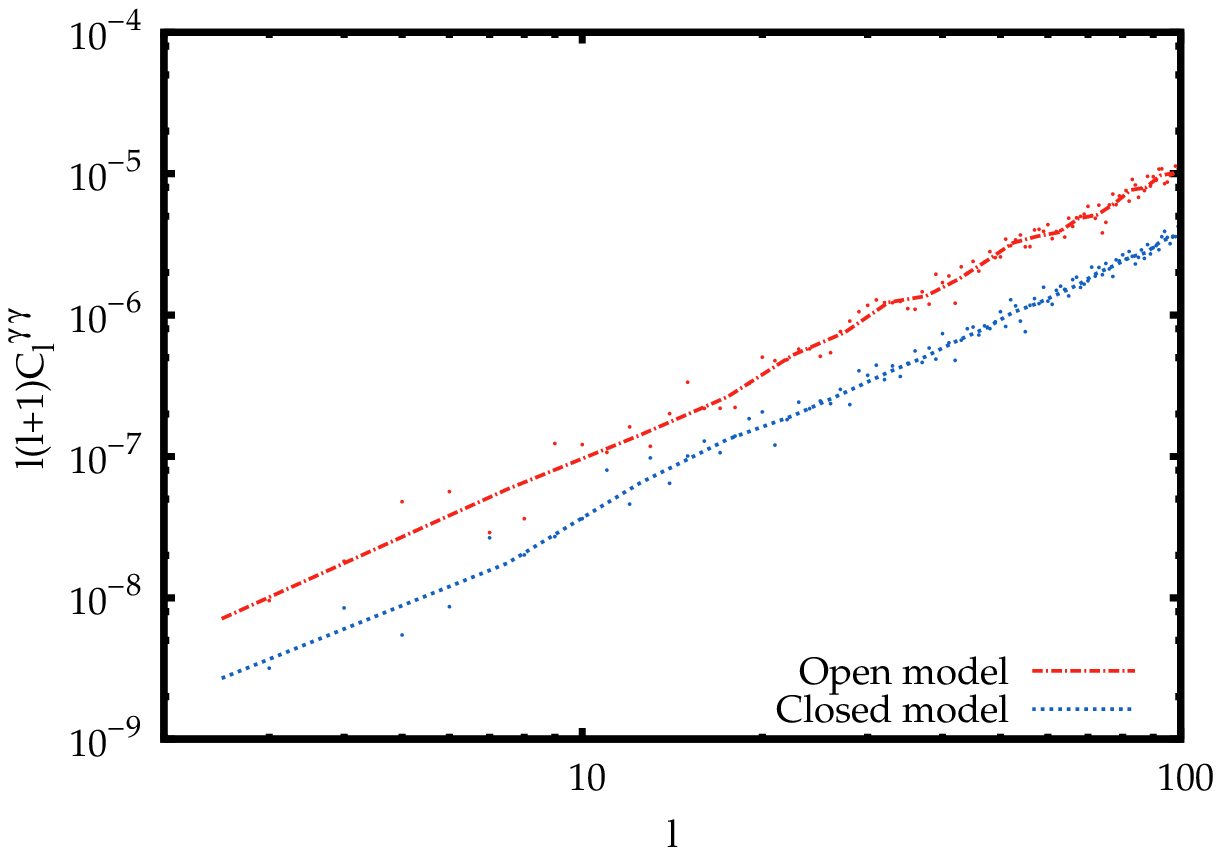} 
\caption{Power spectra for angular diameter distance anisotropy $\Delta D_A /\bar{D}_A$ and shear $\gamma$ calculated from the Swiss Cheese models. The lines are binned spectra and the dots are individual multipoles.}
\label{fig:cls}
\end{figure}

\noindent The power spectra of $\Delta T/\bar{T}$, $\Delta D_A/\bar{D}_A$
and $\gamma$ are plotted in figures \ref{fig:cltt} and
\ref{fig:cls}, with the full linear perturbation theory CMB spectrum
and the linear ISW effect shown for comparison.

The temperature power spectrum for the open model peaks at a multipole
between $10$ and $20$, with an amplitude approximately three
orders of magnitude below the linear theory ISW power spectrum,
and a bit below the Rees--Sciama effect calculated from perturbation theory
and ray-tracing in simulations \cite{Rees:1968, Seljak:1995, Cai:2010}.
At $l=100$ the Swiss Cheese result is two orders of magnitude below
the linear ISW effect.
The closed model spectrum at its maximum is another three
orders of magnitude below the spectrum of the open model.
It does not have a peak in the multipole range we consider.
This is related to the fact that the temperature profile has
less structure, as shown in figure \ref{fig:Tprofiles}.
The closed model only produces a faint hot spot, whereas in the open model,
the contrasts are stronger, with the cold center surrounded by a hot ring.
Also, in the closed case a single ray receives contributions from a
larger number of holes at different distances and angular scales, because
the temperature perturbation due to the holes goes down more slowly
than in the open case, as mentioned above.
The small amplitudes of the power spectra, especially on large scales,
reflect the fact that the observer and sources are located in the
background, and there are almost no correlations between the holes.

The power spectra for the distance and shear are both featureless,
with almost constant amplitude of $C_l^{DD}$ and $C_l^{\gamma\gamma}$.
The amplitude of the distance power spectrum is
$l(l+1)C_l^{DD}(l=100)=2\cdot 10^{-4}$ in the open model,
and a factor of 3 smaller in the closed model.
For the shear power spectrum, we have
$l(l+1)C_l^{\gamma\gamma}(l=100)=1\cdot10^{-5}$ in the open model
and; the closed model result is again down by a factor of 3.
These numbers are in agreement with the values of
$\sigma_{\Delta D_A/\bar{D}_A}$ and $\sigma_{\gamma}$ listed in
table \ref{table:errors}.

\paragraph{Comparison with other studies.}

Let us discuss previous work, some of which has claimed surprisingly
large effects on the distance, despite the average expansion rate
being close to the FRW case. Such results have come down to selection
effects, which were already the focus of the work of Zel'dovich
\cite{Zeldovich:1964}. We can distinguish four different cases of sampling bias.

First, both the distribution of structures and the choice
of lines of sight can violate statistical homogeneity and
isotropy\footnote{Models where the observer is at the centre of
a void whose radius is of the order of the size of the visible
universe are an extreme case
\cite{Mustapha:1998, Enqvist:2007, GarciaBellido:2008, Enqvist:2009, Bolejko:2011b, Sundell:2015}.}.
In \cite{Marra:2007pm, Marra:2008, Bolejko:2011a, Kostov:2009, Kostov:2010},
there were large voids perfectly aligned so that the light
rays pass through their centres, leading to an effect of order unity;
a smaller but still sizeable effect on the distance with aligned voids
was also found in \cite{Clifton:2009a}.
Randomisation of the voids strongly reduces the effect
\cite{Vanderveld:2008, Kostov:2009, Kostov:2010}.

Second, even though the though the distribution of holes may be
statistically homogeneous and properly randomised lines of sight
may be chosen, the distribution of structures can lack statistical
isotropy.
In \cite{Valkenburg:2009} the power spectra of the temperature and
the angular diameter distance were calculated in a Swiss Cheese model
with a cubic lattice of LTB holes up to $z = 1.92$.
The dependence of the signal on the size of the hole was studied,
with radii between $3.5 \ \mathrm{Mpc}$ and $1.75 \ \mathrm{Gpc}$ considered.
The temperature power spectrum due to holes with radius $35 \ \mathrm{Mpc}$ 
was found to be of the same order of magnitude as the primary CMB signal.
This is unexpectedly large, especially as the model does not have a
cosmological constant, so the linear ISW effect vanishes.
A possible reason is the higher regularity of the lattice, as
suggested in \cite{Kostov:2009, Kostov:2010}. The fact that
$t_B\neq0$ in \cite{Valkenburg:2009}, and so there is a decaying
mode, may also enhance the temperature perturbation.
In contrast, the amplitude of the distance power spectrum
on small scales is similar to that in our models, with
$l(l+1)C_l^{DD}(l=100)\sim10^{-4}$, though on large scales
the spectrum in \cite{Valkenburg:2009} is considerably larger
than in our case, because it stays flat whereas our amplitude falls.
The variation for single beams is comparable to our results,
$\sigma_{\Delta D_A/\bar{D}_A}=4.3\cdot10^{-3}$.
We conclude that the distance spectrum appears to be less sensitive to
model details, and there is more room to change the
distance-redshift relation in Swiss Cheese models without having
an unacceptably large effect on the CMB than estimated in
\cite{Valkenburg:2009}, in agreement with \cite{Kostov:2009, Kostov:2010}.

A third form of bias can arise from not sampling all lines
of sight. For example, \cite{Fleury:2013sna, Fleury:2013uqa, Fleury:2014}
found the luminosity distance in a Swiss Cheese model to be of the
order 10\% larger than in the background, using both a hexagonal
and a random arrangement of holes with radii from 1 Mpc to 200 Mpc.
The large deviation is due to the fact that only light rays that
passed sufficiently far from the centres of the Schwarzschild holes
were considered, to model opaque clumps of matter. This leads
to undersampling the density in the shell around the Schwarzschild hole
and a Dyer--Roeder-like distance-redshift relation.
The level of bias in point sources such as supernovae due to opaqueness
in the real universe is not fully settled
\cite{Rasanen:2008be, Bolejko:2010c, Bolejko:2011a, Bolejko:2012ue, Busti:2013, Clarkson:2011br}.

Fourth, even if all lines of sight are considered, such as when studying
the CMB (leaving aside sky cuts due to foregrounds), the result can be
biased if the lines of sight are not correctly weighted.
In \cite{Clarkson:2014pda} it was claimed that there is a
systematic increase in the mean angular diameter distance of
the order $1\%$ from second order perturbation theory (for
calculations of the angular diameter distance in perturbation theory, see
\cite{Sasaki:1987, Sugiura:1999a, Bonvin:2005ps, Vanderveld:2007, BenDayan:2012a, BenDayan:2012b, Umeh:2012, BenDayan:2012c, BenDayan:2013, Fanizza:2013, Marozzi:2014, Umeh:2014, Fanizza:2015}).
However, there was a subtle sampling bias: the average was
taken over the ensemble rather than sky directions, an issue that
can potentially also arise in other contexts where a statistical treatment
of light propagation is used (as opposed to considering one fixed spacetime,
as in our case).
It was argued in \cite{Bonvin:2015kea} that even though the distance
perturbation for a typical light ray is positive, the negative contributions
outweigh the positive ones when averaged over the sky.
Also, one has to make a careful distinction between perturbations
in the flux and in the luminosity distance, as they are not linearly
dependent. With these biases taken into account, the effect is reduced
by orders of magnitude \cite{Bonvin:2015uha, Bonvin:2015kea, Kaiser:2015}
(see also \cite{Fanizza:2015}).

These arguments have been related to the assumption of \cite{Weinberg:1976}
that the area average of the flux (proportional to $\mu$)
is conserved if the area of a sphere is unaffected by inhomogeneities.
It can be shown that in the same situation the angular average of
the inverse flux (proportional to $\mu^{-1}$) is conserved
\cite{Kibble:2004, Kaiser:2015}.
As discussed in the introduction, this is not true in general.
We can now see how well the conservation holds in our Swiss Cheese universe,
whose average properties are by construction close to the FRW case.
We have considered angular averages, in which case it follows from
$\langle\mu^{-1}\rangle=1$ straightforwardly, using \re{mu} and \re{mud}, that
$\langle\Delta D_A/\bar{D}_A\rangle=-\frac{1}{2}\langle\kappa^2\rangle=-\frac{1}{2}\langle(\Delta D_A/\bar{D}_A)^2\rangle$.
In particular, this implies that $\langle\Delta D_A/\bar{D}_A\rangle<0$.
As we have noted, we do not have enough beams to see a statistically
significant deviation of $\langle\Delta D_A/\bar{D}_A\rangle$ from 0,
but we can see whether it is as negative as required
by the above relation. The distribution of
$\langle\Delta D_A/\bar{D}_A\rangle$ is shown in figure \ref{fig:bootstrap},
and some numbers are given in table \ref{table:errors}.
In the closed case, the errors are large, and our results are consistent
with the relation
$\langle\Delta D_A/\bar{D}_A\rangle=-\frac{1}{2}\langle(\Delta D_A/\bar{D}_A)^2\rangle$.
In the open case, the errors are smaller, and the probability that
$\langle\Delta D_A/\bar{D}_A\rangle$ is as small as
$-\frac{1}{2}\langle(\Delta D_A/\bar{D}_A)^2\rangle$ is only 1.4\%.
Another way of viewing the same thing is to look at $\langle\mu^{-1}\rangle-1$,
given in table \ref{table:errors}. In the closed case, zero is within
1$\sigma$, whereas in the open case, the probability for zero is 1.4\%.
We note that, in contrast, $\langle\mu\rangle-1$ is consistent with zero
at the 1$\sigma$ level in the closed case and 2$\sigma$ level in the open
case, and the corresponding relation 
$\langle\Delta D_A/\bar{D}_A\rangle=\frac{3}{2}\langle(\Delta D_A/\bar{D}_A)^2\rangle$
is easily satisfied.
The studies \cite{Flanagan:2011, Peel:2014}, discussed below, also find
$\langle\Delta D_A/\bar{D}_A\rangle>0$.
Therefore, the assumption that the area of spheres is unperturbed,
central to the arguments in \cite{Weinberg:1976, Kibble:2004, Kaiser:2015},
does not seem to be a good approximation in these Swiss Cheese models.
Using more beams would allow to study the issue in our model with
higher statistical significance.

In \cite{Bonvin:2005ps}, the power spectrum for the luminosity distance
was calculated in linear perturbation theory in a $\Lambda$CDM universe
(see also \cite{Sasaki:1987, Sugiura:1999a}).
At multipoles $l>10$ and large redshifts, lensing by structures
is the dominant effect. Because of the distance duality relation (\ref{eq:reci}),
perturbations of the angular diameter and luminosity distance
are comparable,
$\frac{\Delta D_A}{\bar{D}_A}\simeq\frac{\Delta D_L}{\bar{D}_L}$,
as long as the redshift perturbation is much smaller.
Their distance power spectrum has a similar shape as in our case,
and the amplitude $l(l+1)C_l^{DD}(l=100)=3\cdot10^{-5}$ is close to
our result for the closed model. In our open model the amplitude is
an order of magnitude higher.

In \cite{Flanagan:2011}, a $\Lambda$CDM Swiss Cheese model with
spherical voids of 35 Mpc radius was studied with $N=2\cdot10^6$
beams up to $z=2.1$.
For $z=1$, the authors report (translating from the distance modulus
to the angular diameter distance) 
$\langle\Delta D_A/\bar{D}_A\rangle=(1.4\pm0.5)\cdot10^{-3}$.
The variation for individual beams is $\sigma_{\Delta D_A/\bar{D}_A}=0.01$.
The shift in the mean is an order of magnitude larger than in our case,
and the variation for single beams is a factor of a few larger.
One reason may be that in \cite{Flanagan:2011} the time spent
by the light rays in the cheese is minimised. 
In \cite{Flanagan:2012} halos were included, but the resulting
holes are no longer solutions of the Einstein equation, and the
treatment is statistical rather than exact. The results were
$\langle\Delta D_A/\bar{D}_A\rangle=-(5\ldots6)\cdot10^{-4}$ and
$\sigma_{\Delta D_A/\bar{D}_A}=0.03$, an order of magnitude
higher than in our open model. The sign of the mean shift is
also the opposite of that in exact in Swiss Cheese calculations.

In \cite{Peel:2014}, a $\Lambda$CDM Swiss Cheese model with Szekeres holes
that are not spherically symmetric was considered, up to $z=1.5$.
Holes with different radii were studied, including ones with radius
$35 \ \mathrm{Mpc}$, close to our case, and a density profile
somewhat similar to that of our open model.
For $z=1$ the authors report
$\langle\Delta D_A/\bar{D}_A\rangle = 4.5\cdot10^{-4}$ and
$\sigma_{\Delta D_A/\bar{D}_A} = 1.8\cdot10^{-3}$ for sources at $z=1$.
They have $N=1000$, so the error estimate for
$\langle\Delta D_A/\bar{D}_A\rangle$ is $0.6\cdot10^{-4}$,
and the positive mean shift is statistically highly significant.
It is somewhat surprising that their standard deviation
is smaller than in our model, while the mean distance shift is larger.
As noted, we do not find a statistically significant shift in the
mean distance, only a limit of
$|\langle\Delta D_A/\bar{D}_A\rangle|\lesssim10^{-4}$.
Given that the Szekeres holes in \cite{Peel:2014} are less symmetric
than our spherical holes, we would have expected the deviation for
a single beam to be larger \cite{Troxel:2013}, as it takes longer
for the inhomogeneities along a light ray to average out.

Observational constraints on fluctuations in the distance to the CMB
are mainly from studies of the variation of the multipole location
of the acoustic peaks \cite{Hansen:2004vq, Hansen:2008}, which
(for fixed matter content) is a measure of the angular diameter distance
\cite{Vonlanthen:2010, Audren:2013a, Audren:2013b}. The current
precision is below the theoretically expected signal.

\section{Conclusions}

\paragraph{Effect of random holes on the CMB.}

We have done the first calculation of the redshift, distance and
shear of the CMB in a Swiss Cheese model with randomised holes in a
$\Lambda$CDM background. We have been careful about sampling
lines of sight fairly, and have used a spacetime with fixed 
hole positions drawn from a uniform distribution.
We have considered two kinds of LTB holes, with either an
over- or underdensity in the centre, called the open and closed model,
respectively.
The hole radius is $r_b=50 \mpc$, but the profile becomes close
to the FRW cheese at a radius of about 30$\mpc$ in the open and
20$\mpc$ in the closed model.

We find a maximum temperature perturbation for a single hole of
$|\Delta T/\bar{T}|\sim10^{-6}\sim(r_b H_0)^3$ in the open case, for a
hole close to the observer; the amplitude falls off sharply with
the distance to the hole. The result is sensitive to the hole
profile, for the closed case the amplitude is an order of magnitude
smaller and the fall-off is less strong.
The difference in the order of magnitude corresponds, roughly,
to the ratio of the average expansion rate in the open and
closed models, and somewhat more closely to the ratio of the
proper volume average of the time derivative of the density contrast.
The distance perturbation is less sensitive to the details of the hole. For
a single hole the distance perturbation has nearly the same amplitude in
both models, although the dependence on the viewing angle is distinctive.
The maximum amplitude is $\sim10^{-3}$ for holes located
at a distance of $\sim100 r_b$ from the observer, whereas
the typical amplitude is $10^{-4}\sim(r_b H_0)^2$. The amplitude of
the integrated null shear is similar, with $\gamma\sim10^{-4}$ in
both the open and closed case.

For the entire distribution of holes, we have calculated the
the shift in the mean, variation for a single beam and
power spectra for $\Delta T/\bar{T}$, $\Delta D_A/\bar{D}_A$
and $\gamma$, using 12 288 beams and an equal number of pixels on the sky.
The errors are estimated using a bootstrap algorithm.
The sky average of the shift in the temperature is
$\langle\Delta T/\bar{T}\rangle=(2.15\pm0.10)\times10^{-8}$
for the open model and $(3.17\pm0.03)\times10^{-9}$ for the closed model,
a difference of an order of magnitude, as in the single hole case.
For the open model, the power spectrum peaks at a multipole between
10 and 20, but even there it is below the Rees--Sciama effect
estimated from perturbation theory and ray-tracing simulations.
On all the scales we consider, $l\lesssim100$, the open model
power spectrum is two or more orders of magnitude below the linear
theory ISW power spectrum. The closed model power spectrum, in turn,
is at least two orders of magnitude below that of the open model.

For the angular diameter distance, we find 
$\langle\Delta D_A/\bar{D}_A\rangle=(8.5\pm4.3)\cdot10^{-5}$
in the open model and $(1.1\pm2.4)\cdot10^{-5}$ in the closed model.
In other words, we do not see a statistically significant systematic
shift, we can only quote a 95\% limit of
$|\langle\Delta D_A/\bar{D}_A\rangle|\lesssim10^{-4}$.
It is important to account for the variation between
different sky realisations when drawing conclusions about the
mean shift, especially as the typical variation between lines of sight
is $\sim10^{-3}$ for both the open and closed model, much larger than
the possible change in the mean. The amplitude of the power spectrum is
$l(l+1)C_l^{DD}(l=100)=2\cdot 10^{-4}$ in the open model,
and a factor of 3 smaller in the closed model.

For the integrated null shear, we find
$\langle\gamma\rangle=(1.78\pm0.01)\times10^{-3}$ in the open model
and $(9.85\pm0.08)\times10^{-4}$ in the closed model.
Its shift when passing through many holes is larger than the shift
for a single hole, unlike for the temperature and distance.
The amplitude of the power spectrum is
$l(l+1)C_l^{\gamma\gamma}(l=100)=1\cdot10^{-5}$ for the open model,
and the closed model result is again a factor of 3 smaller.

We have compared our results to earlier work, some of which has had
selection biases leading to larger results for the distance or temperature.
The temperature power spectrum is rather sensitive to the details of the
inhomogeneities, with variations of orders of magnitude between
different models. The distance power spectrum is more robust and, taking
selection effects in account, our results are comparable to most previous
Swiss Cheese calculations, as well as results from perturbation theory.
However, some works have reported larger mean shifts in the distance,
up to $\sim10^{-3}$.

We have considered the argument that the areas of spheres are
unaffected by perturbations, leading to conservation of the angular
average of the inverse flux, and the relation
$\langle\Delta D_A/\bar{D}_A\rangle=-\frac{1}{2}\langle(\Delta D_A/\bar{D}_A)^2\rangle$
\cite{Weinberg:1976, Kibble:2004, Kaiser:2015}.
Our closed model is consistent with this relation, whereas in the
open model the probability that the two sides agree is only 1.4\%.
Using more beams than our 12 288 would make it possible to draw more
definite conclusions. 

The CMB null shear has not been calculated using a Swiss Cheese model before.
There is much room for refinement, including correlations among the holes,
changing the packing fraction and using holes of different sizes and
profiles, including non-spherical Szekeres holes.
As lensing of the CMB (and large scale structure) is an increasingly
important cosmological probe, more realistic Swiss Cheese models could
potentially be an interesting way to study it, as they automatically
include all relativistic and non-linear effects,
without the need for perturbation theory.

\acknowledgments

We thank Pierre Fleury, Austin Peel and Sebastian J. Szybka for correspondence.

\appendix

\section{Integrated null shear proof} 
\label{sec:proof}

Let us show that when the null shear $\tsig$ is small,
the integrated null shear $\gamma$ does not depend on
whether the initial conditions are set at the observer
or at the source.
If the null shear is small, we can neglect the term $\tsig^2$
in the distance equation \re{sachs-d}. In that case
\re{sachs-d} becomes independent of the shear equation
\re{sachs-shear}, which correspondingly becomes linear.
The resulting equations can be written as
\bea
  \label{S} \frac{\D^2 S}{\D \l^2} + Q S = 0 \\ 
  \label{ts} \frac{\D \tsig}{\D \l} + \tthe \tsig = f \ , 
\eea

\noindent where we have denoted the distance by $S$,
$Q\equiv\frac{1}{2} R_{\mu \nu} k^\mu k^\nu$,
$\tthe=2 \frac{1}{S}\frac{\D S}{\D \l}$
and $f$ is the Weyl tensor source term.
There are no indices in \re{ts}, as $\tsig$ can be
considered a complex variable that contains
both of the independent degrees of freedom of $\tsig_{\mu\nu}$
(see e.g. \cite{P.Schneider1992}, p. 106),
in which case $f$ is also complex.  Equivalently,
we could write \re{ts} separately for $\tsig_1$ and $\tsig_2$.

We have two different initial conditions for the equations \re{S}, \re{ts}.
We can either start from the observer, so that, as in \re{init},
\bea \left. S \right|_{O} = 0, \hspace{0.5cm} \left. \frac{\D S}{\D \l}\right|_{O} = -H_0^{-1}, \hspace{0.5cm} \left.\tsig\right|_{O} = 0 \ ,
\eea

\noindent or from the source, in which case we have
\bea \left. S \right|_{S} = 0, \hspace{0.5cm} \left. \frac{\D S}{\D \l}\right|_{S} = A, \hspace{0.5cm} \left.\tsig\right|_{S} = 0 \ ,
\eea

\noindent where $A>0$ is a constant. (Note that setting the initial
conditions for $S$ and $\tsig$ at different ends would be inconsistent.)
We denote the two cases by the subscripts $O$ and $S$, so we have four
functions, $S_O(\l), S_S(\l), \tsig_O(\l)$ and $\tsig_S(\l)$.

From \re{ts} we get the solution
\bea
  \tsig_O = S_O(\l)^{-2} \int_{\l_O}^{\l}\D\l' S_O(\l')^{2} f(\l') \ ,
\eea

\noindent and correspondingly for $\tsig_S$ with the substitution
$O\rightarrow S$. We define
\bea
  \label{go} \gamma_O &\equiv& \int_{\l_O}^{\l_S} \D\l \tsig_O(\l) = \int_{\l_O}^{\l_S}\D\l S_O(\l)^{-2} \int_{\l_O}^{\l}\D\l' S_O(\l')^{2} f(\l') \\
  \label{gs} \gamma_S &\equiv& \int_{\l_S}^{\l_O} \D\l \tsig_S(\l) = \int_{\l_S}^{\l_O}\D\l S_S(\l)^{-2} \int_{\l_S}^{\l}\D\l' S_S(\l')^{2} f(\l') \ .
\eea

It is well known that $S_O=D_A$ and that $S_S$ gives $(1+z)D_A$
\cite{Eth33,Ellis:1971pg}. Let us now show that the integrated
null shear satisfies $\gamma_O=\gamma_S$
(in the case when $\tsig$ small, so that \re{S} applies).
As $S_O$ and $S_S$ are solutions of the same differential equation \re{S},
they are related as \cite{Rosquist:1988}\footnote{We have a different
convention for the sign and normalisation of the affine parameter than
\cite{Rosquist:1988}.}
\bea
  \label{So} S_O(\l) &=& C_O S_S(\l) \int_{\l_O}^{\l}\D\l' S_S(\l')^{-2} \\
  \label{Ss} S_S(\l) &=& C_S S_O(\l) \int_{\l_S}^{\l}\D\l' S_O(\l')^{-2} \ ,
\eea

\noindent where  $C_O=-S_O(\l_S) \frac{\D S_S}{\D \l}(\l_S)$ and
$C_S=-S_S(\l_O) \frac{\D S_O}{\D \l}(\l_O)$. Taking a derivative
of either \re{So} or \re{Ss} shows that $C_O=-C_S$. Partial integration
of \re{go} gives
\bea
  \gamma_O &=& - \int_{\l_O}^{\l_S}\D\l \left[ \int_{\l_S}^{\l}\D\l' S_O(\l')^{-2} \right] S_O(\l)^{2} f(\l) \nonumber \\
  &=& - C_S^{-1} \int_{\l_O}^{\l_S}\D\l S_O(\l) S_S(\l) f(\l) \ ,
\eea

\noindent where we have on the second line applied \re{Ss}. Repeating
the exercise with $\gamma_S$ gives the same result, with the change
$O\leftrightarrow S$. Given that $C_O=-C_S$ and changing the direction
of integration also gives a minus sign, we have the desired equality
$\gamma_O=\gamma_S$.

This result also follows from the reciprocity relation for the Jacobi matrix (eq. (35) of \cite{Perlick:2010})\footnote{We thank Pierre Fleury for pointing this out.}.
The reciprocity relation is not limited to the case when $\tsig$ is small, but if
$\tsig$ is large, its relation of the lensing parameters of the Jacobi matrix is more
complicated, and $\gamma_O=\gamma_S$ does not hold, given the definitions
\re{go} and \re{gs}.

\bibliographystyle{JHEP2}

\bibliography{refs}

\end{document}